\pdfoutput=1
\documentclass[12pt,a4paper]{amsart}
\usepackage[pdftex]{graphicx}
\usepackage{amssymb}
\usepackage{amsmath}
\usepackage{subfigure}
\usepackage[ruled,linesnumbered,norelsize]{algorithm2e}
\renewcommand{\algorithmcfname}{ALGORITHM}
\SetAlFnt{\small}
\SetAlCapFnt{\small}
\SetAlCapNameFnt{\small}
\SetAlCapHSkip{0pt}
\IncMargin{-\parindent}
\usepackage[numbers]{natbib}
\usepackage[all]{xy}
\usepackage{subfigure}
\usepackage{url}

\providecommand{\set}[2]{\left\{#1 \mid #2\right\}}
\providecommand{\propagate}[2]{\mathrm{propagate}\left(#1,#2\right)}
\providecommand{\analyze}[2]{\mathrm{analyze}\left(#1,#2\right)}
\providecommand{\resolve}[4]{\mathrm{resolve}\left(#1,#2,#3,#4\right)}
\providecommand{\resolveplus}[6]{\mathrm{resolveplus}\left(#1,#2,#3,#4,#5,#6\right)}
\providecommand{\extendobdd}[3]{\mathrm{extendobdd}\left(#1,#2,#3\right)}
\providecommand{\makeformula}[3]{\mathrm{makeformula}\left(#1,#2,#3\right)}
\providecommand{\associate}[5]{\mathrm{associate}\left(#1,#2,#3,#4,#5\right)}
\providecommand{\backtrack}[2]{\mathrm{backtrack}\left(#1,#2\right)}
\providecommand{\intv}[2]{[#1,#2]}
\providecommand{\intvr}[2]{[#1,#2)}
\providecommand{\intvlr}[2]{(#1,#2)}

\newtheorem{proposition}{Proposition}

\newtheorem{definition}{Definition}
\newtheorem{example}{Example}

\newtheorem{remark}{Remark}

\title{Implementing Efficient All Solutions SAT Solvers}

\author[T.\ Toda]{Takahisa Toda}
\address[T.\ Toda]{Graduate School of Information Systems, University of Electro-Communications, 1-5-1 Chofugaoka, Chofu, Tokyo 182-8585, Japan.}
\email{takahisa.toda@is.uec.ac.jp}

\author[T.\ Soh]{Takehide Soh}
\address[T.\ Soh]{Information Science and Technology Center, Kobe University, 1-1 Rokkodai, Nada Kobe 657-8501, Japan.}
\email{soh@lion.kobe-u.ac.jp}

\begin{document}

\maketitle

\begin{abstract}
All solutions SAT (AllSAT for short) is a variant of propositional satisfiability problem.
Despite its significance, AllSAT has been relatively unexplored compared to other variants.
We thus survey and discuss major techniques of AllSAT solvers.
We faithfully implement them and conduct comprehensive experiments using a large number of instances and various types of solvers including one of the few public softwares.
The experiments reveal solver's characteristics.
Our implemented solvers are made publicly available so that other researchers can easily develop their solver by modifying our codes and compare it with existing methods.
\end{abstract}

\section{Introduction}\label{sec:intro}
\emph{Propositional satisfiability} (\emph{SAT} for short) is to decide if a Boolean formula is satisfiable.
SAT is ubiquitous in computer science.
Because of its significance, it has attracted the attention of many researchers from theory to practice.
Many applications have motivated empirical studies, in particular the development of \emph{SAT solvers}, softwares to solve satisfiability.
A fundamental task of SAT solvers is to solve as many instances as possible in a realistic amount of time.
To this end, various practical algorithms and elegant implementation techniques have been developed~\cite{vanHarmelen:2007:HKR:1557461}~\cite{Biere:2009:HSV:1550723}~\cite{Malik:2009:BST:1536616.1536637}.

There are many variants of SAT.
\emph{All solutions SAT} (\emph{AllSAT} for short) or \emph{model enumeration} is studied in the paper.
It is, given a CNF formula, to generate partial satisfying assignments such that they form a logically equivalent DNF formula.
Compared to neighboring areas, AllSAT has been relatively unexplored.
This is mentioned in the literature~\cite{DBLP:conf/iri/JabbourLSS14} and also supported by the fact that there are only a few recent papers, almost no software\footnote{A few exceptions are clasp~\cite{clasp}, picosat~\cite{DBLP:journals/jsat/Biere08}, and relsat~\cite{Bayardo:1997:UCL:1867406.1867438}. Although they support solution generation, they are positioned as an answer set solver, a single solution SAT solver, and a \#SAT solver rather than AllSAT solvers, respectively.} is publicly available, and it has not even been taken up in major handbooks related to satisfiability.


A recent application of AllSAT is data mining.
A fundamental task in data mining is to generate all interesting patterns from a given database~\cite{1384-5810}. 
Examples include frequent itemsets, maximal frequent itemsets, or closed itemsets in transaction databases.
Although algorithms for generating various patterns have been proposed, they are basically specialized for their target patterns.
This means that different patterns require new algorithms.
For this reason, a framework based on declarative paradigm has recently been proposed~\cite{Guns20111951}.
A basic flow is that constrains of patterns to be generated are formulated as logical formulae and solved with a generic solver.
Hence, all that users do is simply to model their problems, not to design algorithms.
Among much related work, an approach on which problems are encoded into CNF formulae and solved with AllSAT solvers has been studied~\cite{Jabbour:2013:BSS:2505515.2505577}.
An advantage of declarative paradigm is its ability to handle new patterns in a flexible manner.
There is no need to see details of algorithms on which solvers are based, thereby it is opened to wider users.
It is instead inferior in efficiency to problem-specific approaches.
In practice it is necessary to balance between efficiency and flexibility.
Therefore, improving solver's performance is essential in the declarative framework.

Besides the data mining, there have been many studies on the application of ALLSAT, in particular to formal verification, such as network verification~\cite{978-3-642-33386-6_1}~\cite{export:201589}~\cite{6873212}, predicate abstraction~\cite{0925-9856}, backbone computation~\cite{Marques-Silva:2010:CBP:1860967.1860972}, image and preimage computation in unbounded model checking~\cite{978-3-540-41219-9}~\cite{McMillan}~\cite{Sheng:2003:EPC:789083.1022826}~\cite{DBLP:conf/date/LiHS04}~\cite{1382631}~\cite{grumberg}.

Considering the above, we find it important to clarify state-of-the-art techniques of AllSAT solvers and to improve them on a firm basis.
However, there are the following issues in the existing researches of ALLSAT.
\begin{itemize}
\item Several methods are proposed but they are not globally compared. It is thus difficult to decide which method is effective for which kinds of ALLSAT instances.
\item Experiments are not carried out on comprehensive benchmarks.
\item There is few public ALLSAT solver, which makes it difficult to compare existing techniques.
\end{itemize}
We thus would like to survey major techniques of AllSAT solvers and try to complement past references by gathering and organizing existing techniques.
We further add some novel techniques.
To evaluate solvers, we conduct experimental comparisons, including \emph{clasp}, one of the few off-the-shelf softwares with solution generation support.
Our implemented solvers are made publicly available with expectation that further improvement on solvers and their evaluation are easily done and the AllSAT research is stimulated.

The paper is organized as follows.
Section~\ref{sec:related} provides related work of AllSAT.
Section~\ref{sec:pre} provides necessary notions, terminology, and results.
Section~\ref{sec:tech} surveys major techniques of AllSAT solvers, where those including our original ideas are indicated by adding asterisks to their titles.
Section~\ref{sec:exp} provides experimental results.
Section~\ref{sec:conc} concludes the paper.

\section{Related Work}\label{sec:related}
Another variant of SAT is the \emph{dualization} of Boolean functions: given a DNF formula of a Boolean function $f$, it is to compute the complete DNF formula of the dual function $f^{d}$.
Since a CNF formula of $f^{d}$ can be easily obtained by interchanging logical disjunction with logical conjunction as well as the constants $0$ with $1$, a main part is to convert CNF to the complete DNF.
Hence, an essential difference from AllSAT is that the resulting DNF formula must be complete.
Dualization has been well-studied in terms of complexity~\cite{Eiter20082035}~\cite{crama2011boolean}, while there seems no recent empirical study with a few exceptions~\cite{Jin:2005:PCF:1065579.1065775}~\cite{978-3-642-22109-5}~\cite{primegen}~\cite{DBLP:conf/isaim/Toda14}.
Practical algorithms for a restricted form of dualization have been presented~\cite{todasea}~\cite{Murakami201483} and some implementations are available\footnote{Hypergraph Dualization Repository, by Keisuke Murakami and Takeaki Uno, \url{http://research.nii.ac.jp/~uno/dualization.html}, accessed on 19th Jan., 2013. HTC-DD: Hypergraph Transversal Computation with Binary Decision Diagrams, by Takahisa Toda at ERATO MINATO Discrete Structure Manipulation System Project, Japan Science and Technology Agency, at Hokkaido University, \url{http://www.sd.is.uec.ac.jp/toda/htcbdd.html}, accessed on 10th Sept., 2015.}, though they are not for arbitrary Boolean functions.

Another variant is the problem of counting the number of total satisfying assignments, called \emph{propositional model counting} or \emph{\#SAT}.
It has been well-studied because of good applications such as probabilistic inference problems and hard combinatorial problems, and some solvers are available~\cite{Biere:2009:HSV:1550723}.
Although \#SAT is apparently similar to AllSAT, techniques such as connected components and component caching are inherent in counting, and they are not applicable to AllSAT as is~\cite{1562927}.

\section{Preliminaries}\label{sec:pre}
Necessary notions, terminology, and results concerning Boolean functions, satisfiability solvers, and binary decision diagrams are presented in this section.

\subsection{Boolean Basics}\label{subsec:boolean}
A \emph{literal} is a Boolean variable or its negation.
A \emph{clause} is a finite disjunction of literals, and a \emph{term} is a finite conjunction of literals.
A propositional formula is in \emph{conjunctive normal form} (\emph{CNF} for short) if it is a finite conjunction of clauses and in a \emph{disjunction normal form} (\emph{DNF} for short) if it is a finite disjunction of terms.
We identify clauses with sets of literals and CNF formulae with sets of clauses.
The same applies to terms and DNFs.
The \emph{dual} of a Boolean function $f$ is the function $f^d$ defined by $f^d(x_1,\ldots,x_n)=\lnot f (\lnot x_1,\ldots, \lnot x_n)$.
An \emph{implicant} of a Boolean function $f$ is a term $t$ with $t\leq f$, where $t$ is considered as a Boolean function and the order of Boolean functions is introduced as $t\leq f$ if $t(v)\leq f(v)$ for all $v\in\{0,1\}^{n}$.
An implicant is \emph{prime} if the removal of any literal results in a non-implicant.
A DNF formula is \emph{complete} if it consists of all prime implicants.

An \emph{assignment} to a set $V$ of Boolean variables is a partial function from $V$ to $\{0,1\}$.
A \emph{satisfying} assignment for a CNF formula is an assignment $\nu$ such that the CNF formula evaluates to $1$.
An assignment to $V$ is \emph{total} (or \emph{complete}) if it is a total function, that is, all variables in $V$ are assigned values.
A Boolean formula is \emph{satisfiable} if it has a satisfying assignment.
For simplicity, we say that a literal is assigned a value $v$ if the assignment to the underlying variable makes the literal evaluate to $v$.
If there is no fear of confusion, we identify an assignment function $\nu$ over $V$ with the set of the form $\set{(x,v)\in V\times\{0,1\}}{\nu(x)=v}$.
We further identify the assignment $x\mapsto v$ with $x$ if $v=1$ and with $\lnot x$ if $v=0$.
In this way, \emph{assignments and literals are used interchangeably} throughout the paper.

\begin{example}
Consider the sequence of literals $x_1,\lnot x_5, x_3$.
This means that $x_1, x_5,x_3$ are selected in this order and the values $1,0,1$ are assigned to them, respectively.
\end{example}

\subsection{Satisfiability Solvers}\label{subsec:satsolvers}
\emph{Propositional satisfiability problem} (\emph{SAT} for short) is the problem of deciding if there exists a satisfying assignment for a CNF formula.
Algorithm~\ref{alg:cdcl} shows a basic framework on which modern SAT solvers are based.
For simplicity, other techniques such as lazy data structures, variable selection heuristics, restarting, deletion policy of learnt clauses are omitted.
See for details~\cite{vanHarmelen:2007:HKR:1557461} and~\cite{Biere:2009:HSV:1550723}.

A basic behavior of Algorithm~\ref{alg:cdcl} is to search a satisfying assignment in such a way that a solver finds a candidate assignment by assigning values to variables and if the assignment turns out to be unsatisfying, the solver proceeds to the next candidate by backtracking.
The extension of an assignment is triggered at the decide stage, where an unassigned variable $x$ is selected and a value $v\in\{0,1\}$ is assigned to $x$, and it is then spread at the deduce stage, where assignments to other variables are deduced from the most recent decision $(x,v)$.

\emph{Decision assignments} are those given at the decide stage and \emph{decision variables} are those assigned values there.
Consider a decision tree that branches at each decision assignment.
A \emph{decision level} is the depth of that decision tree, which is maintained by the variable $dl$ in Algorithm~\ref{alg:cdcl}.
The decision level of a variable $x$, denoted by $\delta(x)$, is one at which $x$ was assigned a value.
For a literal $l$, the notation $\delta(l)$ is defined as that of the underlying variable, and $l@d$ denotes that $l$, seen as an assignment, was given at level $d$, i.e.\ $\delta(l)=d$.

The deduce stage is described below.
A clause is \emph{unit} if all but one of literals are assigned the value $0$ and the remaining one is unassigned.
The remaining literal is called a \emph{unit literal}.
Unit clauses are important in the deduce stage because assignments to the underlying variables of unit literals are necessarily determined so that unit literals evaluate to $1$.
After some assignments are determined by unit clauses, non-unit clauses may become unit.
Hence, all implications are deduced until an unsatisfied clause exists or no unit clause exists.
This process is called \emph{unit propagation}.
The function $propagate$ performs unit propagation.
\emph{Implied assignments} are those given at this stage and \emph{implied variables} are those assigned values there.
The decision level of an implied variable $x$, the notations $\delta(x)$ and $l@d$, where $l$ is a literal representing an implied assignment, are defined in the same way.

\begin{algorithm}[t]
\SetAlgoNoLine
\KwIn{a CNF formula $\psi$, an empty assignment $\nu$.}
\KwOut{SAT if $\psi$ is satisfied; UNSAT, otherwise.}

$dl              \leftarrow 0$\tcc*[r]{Decision level}

\While{true}{
    $\nu \leftarrow \propagate{\psi}{\nu}$\tcc*[r]{Deduce stage}
    \eIf{conflict happens}
    {
        \lIf{$dl \leq 0$}{return UNSAT}
        $bl \leftarrow \analyze{\psi}{\nu}$\tcc*[r]{Diagnose stage}
        $\nu\leftarrow\set{(x,v)\in\nu}{\delta(x) \leq bl}$\;
        $dl       \leftarrow bl$\;
    }
    {
        \eIf{all variables are assigned values}
        {
            report $\nu$\;
            return SAT;
        }
        {
            $dl          \leftarrow dl+1$\tcc*[r]{Decide stage}
            select an unassigned variable $x$ and a value $v$\;
            $\nu      \leftarrow \nu\cup\{(x,v)\}$\;
        }
    }
}
\caption{DPLL procedure with conflict driven clause learning, where $\delta(x)$ denotes the decision level of a variable $x$.}
\label{alg:cdcl}
\end{algorithm}

\begin{example}\label{ex:cnf}
Consider the CNF formula $\psi$ that consists of the following clauses.
\begin{eqnarray*}
C_1 & = & x_1 \lor \lnot x_3\\
C_2 & = & x_2 \lor x_3 \lor x_5\\
C_3 & = & \lnot x_1 \lor \lnot x_3 \lor x_4\\
C_4 & = & x_4 \lor \lnot x_5 \lor x_6\\
C_5 & = & x_5 \lor \lnot x_6\\
\end{eqnarray*}
Assume the decision assignment $\lnot x_5@1$.
The implied assignment $\lnot x_6@1$ is obtained from $C_5$.
Assume the decision assignment $x_3@2$.
The implied assignments $x_1@2$ and $x_4@2$ are obtained from $C_1$ and $C_3$ in this order.
Assume the decision assignment $x_2@3$, and the CNF formula is satisfied.
\end{example}

It should be noted that in the middle of unit propagation, we may encounter with an unsatisfied clause.
This case is called \emph{conflict}.
As soon as conflict happens, unit propagation halts, even though unit clauses still remain.
In conflict case, if all assigned variables are those assigned prior to any decision (i.e.\ $dl=0$), it means that there are no other assignments to be examined, thereby a CNF formula must be unsatisfiable.
In that case, a solver halts, reporting UNSAT.
If decision has been made at least once, we enter into the diagnose stage to resolve conflict.

At the diagnose stage, a "cause" of the conflict we have just met is analyzed, a new clause is learnt as a result, and it is added to a CNF formula, by which a solver is guided not to fall into the conflict again (and other conflicts related to it).
To do this efficiently, modern solvers maintain an \emph{implication graph} during search, which represents an implication relation between assignments over unit propagation.
Specifically, an implication graph is a directed acyclic graph $G=(V,A)$ such that 
\begin{itemize}
\item vertices in $V$ correspond to literals $l$ representing assignments to their variables;
\item arcs in $A$ correspond to implications so that if a unit clause $C$ with unit literal $l$ yields in unit propagation, then arcs from all assignments to underlying variables in $C\setminus\{l\}$ to the implied assignment $l$ are added;
\item if an unsatisfied clause exists, then arcs from all assignments to underlying variables in that clause to the special vertex $\kappa$ are added.
\end{itemize}
An implication graph might be implemented so that whenever a variable $x$ is implied, it is associated with the clause that determined the assignment to $x$ as a unit clause.
This clause is called the \emph{antecedent} of $x$.

\begin{figure*}[t]
\begin{center}
\includegraphics[width=10cm]{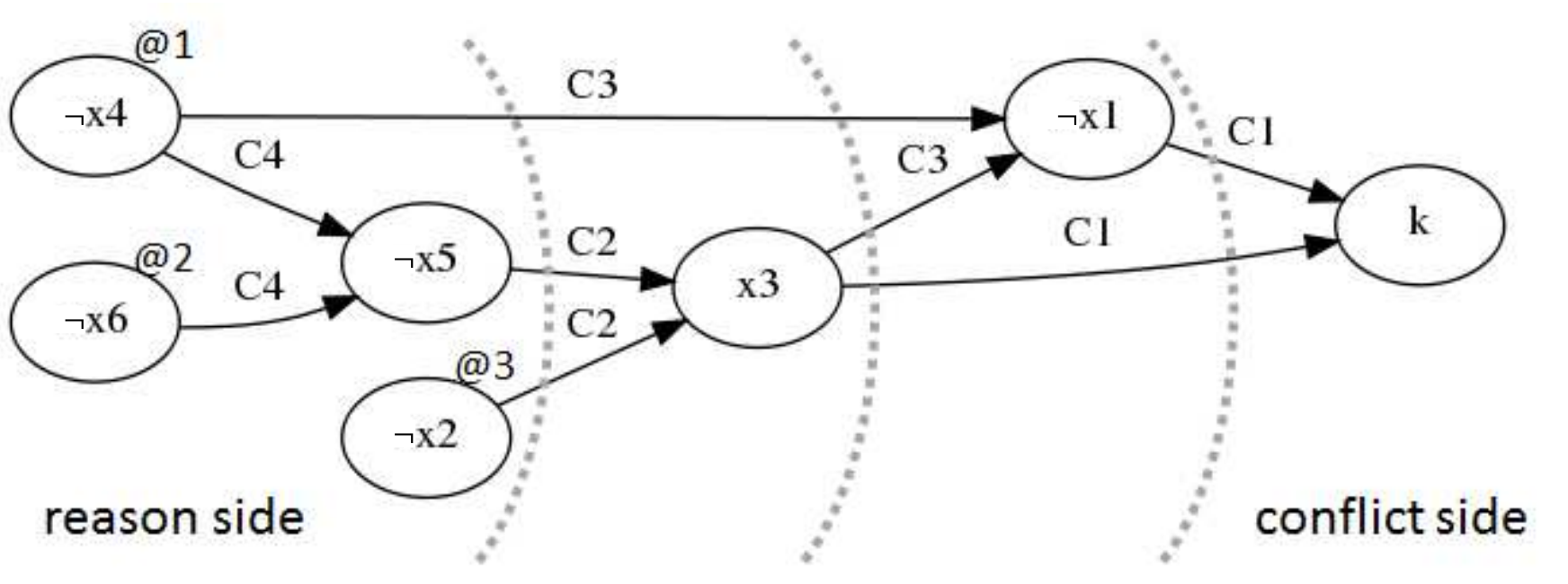}
\end{center}
\caption{A conflict graph, where arcs are labeled with antecedents of their target vertices.}\label{fig:conflictgraph0}
\end{figure*}

\begin{example}
Consider the CNF formula $\psi$ given in Example~\ref{ex:cnf}.
Assume in turn the decision assignments $\lnot x_4@1$, $\lnot x_6@2$, and $\lnot x_2@3$ in this order.
The resulting implication graph is shown in Fig.~\ref{fig:conflictgraph0}.
This case results in conflict because $C_1$ becomes unsatisfied.
\end{example}

A \emph{conflict graph} is a subgraph $H$ of an implication graph $G$ obtained by restricting $G$ so that all vertices have paths to $\kappa$.
For a subset $U$ of vertices in a conflict graph $H=(V,A)$, the \emph{arc-cut} (hereafter \emph{cut}) corresponding to $U$ is the set of arcs that connect vertices in $U$ with those in $V\setminus U$.
Examples are illustrated by dotted curves in Fig.~\ref{fig:conflictgraph0}.

We are now ready to describe a clause learning scheme, which is performed by the function $analyze$.
Consider cuts such that all decision assignments are on one side, called the \emph{reason side}, and the special vertex $\kappa$ is on the other side, called the \emph{conflict side}.
Take negation of literals on reason side that are incident to arcs in a cut, which form a \emph{conflict-driven clause} (or \emph{conflict clause}).
This clause is considered as a "cause" of conflict.
Indeed,if variables are assigned values following literals on reason side that are incident to arcs in a cut, then the same implications on conflict side are derived and the same conflict must take place.
Therefore, to avoid the conflict, it is necessary for variables to be assigned values so that at least one of those literals is negated.
This condition is formulated as the conflict clause obtained above.

As illustrated in Fig.~\ref{fig:conflictgraph0}, there are many choices for cuts that induce conflict clauses.
Among them, conflict clauses that contain exactly one literal from the current decision level are known to be effective.
A \emph{unique implication point} (\emph{UIP} for short) is a vertex in a conflict graph such that every path from the decision at the current decision level to $\kappa$ passes through it.
Note that at least one UIP exists, because the decision at the current decision level is a UIP.
The \emph{first UIP} scheme is to find the UIP that is the closest to $\kappa$.

\begin{example}
Consider the conflict graph given in Fig.~\ref{fig:conflictgraph0}.
The middle curve gives the first UIP $x_3$.
Hence, a conflict clause is $x_4 \lor \lnot x_3$.
\end{example}

The first UIP scheme can be efficiently performed by traversing a conflict graph from $\kappa$ in the reverse order of implications, based on the implementation of an implication graph stated above.
During the traversal, it is easy to decide if the current literal is a UIP.
Consider the cut that induces a UIP.
For all literals on the reason side, their assignments are determined prior to those on the conflict side.
Since the traversal is in the reverse order of implications, the UIP can then be located by keeping track of the number of unvisited vertices that are immediate neighbors of visited vertices.

All that remains is to decide a backtrack level $bl$, cancel all assignments above $bl$, and set $bl$ to the current decision level.
To decide a backtracking level, there are two choices.
Suppose that the decision $p$ at level $i$ resulted in conflict and there is no solution that extends the current assignment.
\emph{Chronological backtracking} cancels all assignments of level $i$ including the decision $p$ and attempts to find a solution by extending the assignment from level $i-1$ with a conflict clause.
In this way, chronological backtracking undoes assignments from a higher to a lower level.
A drawback is that if there is no solution in higher levels, it is hard to get out of those levels.
On the other hand, \emph{non-chronological backtracking} jumps at once to a lower level $j$ by canceling all assignments above $j$ and attempts to find a solution upward by extending assignment from level $j$.
The backtracking level $j$ is commonly determined as the largest level of a variable in a conflict clause below the current level $i$.
For each canceled assignment between the levels $j$ and $i$, the possibility for becoming a solution is left in general, though this does not mean that a solver loses an opportunity to find solutions.

\begin{example}
Consider the conflict graph given in Fig.~\ref{fig:conflictgraph0}.
Since a conflict clause is $x_4 \lor \lnot x_3$, the non-chronological backtrack level is $1$.
Hence, the assignments $\lnot x_6@2,\lnot x_5@2$ are canceled and search restarts from $\lnot x_4@1$, however a solution could be obtained by backtracking to the level $2$.
\end{example}

\begin{remark}
When a conflict clause is learnt, an assignment to its unit literal is implied.
Hence, the function $analyze$ adds the implied assignment to the current assignment function $\nu$, by which its effect to other assignments is taken at a subsequent propagation.
\end{remark}

\subsection{Binary Decision Diagrams}\label{subsec:bdds}
A \emph{binary decision diagram} (\emph{BDD} for short) is a graphical representation of Boolean functions in a compressed form~\cite{BLTJ:BLTJ1585}~\cite{Akers:1978:BDD:1310167.1310815}~\cite{1676819}.
We follow the notation and terminology in Knuth's book~\cite{Knuth:2009:ACP:1593023}.

\begin{figure*}[t]
\begin{center}
        $
        \SelectTips{cm}{}
        \xymatrix@C=0.8pc@R=0.8pc{
        && *+[o][F]{1}\ar@{.>}@/_1pc/[dd]\ar[d] &\\
        && *+[o][F]{2}\ar@{.>}@/^1pc/[ddr]\ar[d]&\\
        && *+[o][F]{3}\ar@{.>}[dl]\ar[dr]&\\
        &*+[F]{\bot} & & *+[F]{\top}\\
        }$
    \caption{The BDD representation for the function: $f(x_1,x_2,x_3)=x_1\land\lnot x_2\lor x_3$.}\label{fig:bdd}
\end{center}
\end{figure*}
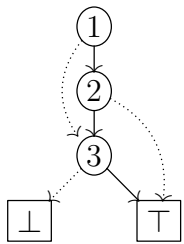

Figure~\ref{fig:bdd} shows an example of a BDD.
Exactly one node has indegree $0$, which is called the \emph{root}.
Each \emph{branch node} $f$ has a label and two children.
Node labels are taken from variable indices, and the children consists of the \emph{LO child} and the \emph{HI child}.
The arc to a LO child is called LO \emph{arc}, illustrated by a dotted arrow, and the LO arc of $f$ means assigning the value $0$ to the variable of $f$.
Similarly, the arc to a HI child is called HI \emph{arc}, illustrated by a solid arrow, and the HI arc of $f$ in turn means assigning the value $1$ to its variable.
There are two \emph{sink nodes}, denoted by $\top$ and $\bot$.
Paths from the root to $\top$ and $\bot$ mean satisfying and unsatisfying assignments, respectively.

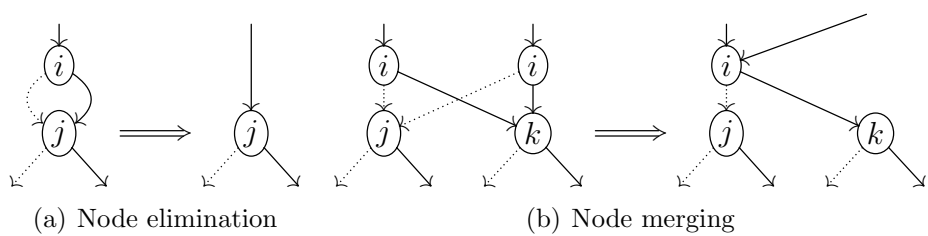
\begin{figure*}[t]
\begin{center}
\subfigure[Node elimination]{$
\SelectTips{cm}{}
\xymatrix @C=0.8pc @R=0.8pc{
&\ar[d]&&&&\ar[dd]&\\
&*+[o][F]{i}\ar@{.>}@/_1pc/[d]\ar@/^1pc/[d]&&&&&\\ 
&*+[o][F]{j}\ar@{.>}[dl]\ar[dr]&\ar@{=>}[rr]&&&*+[o][F]{j}\ar@{.>}[dl]\ar[dr]&\\
&&&&&&\\
}
$\label{fig:bdd_elimination}}
\subfigure[Node merging]{$
\SelectTips{cm}{}
\xymatrix @C=0.8pc @R=0.8pc{
&\ar[d]&&&\ar[d]&&&&\ar[d]&&&\ar[dlll]&\\
&*+[o][F]{i}\ar@{.>}[d]\ar[drrr]&&&*+[o][F]{i}\ar@{.>}[dlll]\ar[d]&&&&*+[o][F]{i}\ar@{.>}[d]\ar[drrr]&&&&\\ 
&*+[o][F]{j}\ar@{.>}[dl]\ar[dr]&&&*+[o][F]{k}\ar@{.>}[dl]\ar[dr]&\ar@{=>}[rr]&&&*+[o][F]{j}\ar@{.>}[dl]\ar[dr]&&&*+[o][F]{k}\ar@{.>}[dl]\ar[dr]&\\
&&&&&&&&&&&&\\
}
$\label{fig:bdd_sharing}}
\caption{The reduction rules}\label{fig:bddrule}
\end{center}
\end{figure*}

BDDs are called \emph{ordered} if for any node $u$ with a branch node $v$ as its child, the index of $u$ is less than that of $v$.
BDDs are called \emph{reduced} if the following reduction operations can not be applied further.
\begin{enumerate}
\item If there is a branch node $u$ whose arcs both point to $v$, then redirect all the incoming arcs of $u$ to $v$, and then eliminate $u$ (Fig.~\ref{fig:bdd_elimination}).
\item If there are two branch nodes $u$ and $v$ such that the subgraphs rooted by them are equivalent, then merge them (Fig.~\ref{fig:bdd_sharing}).
\end{enumerate}
In this paper, ordered reduced BDDs are simply called \emph{BDDs}.
Ordered BDDs that need not be fully reduced are distinguished from ordinary BDDs by calling \emph{OBDDs}.
Note that each node in a BDD (or an OBDD) is conventionally identified with the subgraph rooted by it, which also forms BDD (or OBDD).

\section{Techniques of All Solutions SAT Solvers}\label{sec:tech}
In order to implement an efficient AllSAT solver, we have to carefully determine an appropriate suite of techniques by considering various factors and their characteristics.
Some details have been mentioned only partly and scattered in past references.
We survey major existing techniques of AllSAT solvers and try to complement past references by gathering and organizing existing techniques.
We further add some novel techniques.

This section is organized as follows.
Three major types of solvers are presented in their own subsections.
Each subsection starts with an overview, then provides specific techniques, and ends with the configuration of our implemented solvers.
\emph{We added an asterisk to the title of each specific technique that contains our original ideas.}

\subsection{Blocking Solvers}\label{subsec:bcbased}

\subsubsection{Overview}\label{subsubsec:basicform}
One of the easiest ways of implementing an AllSAT solver is to repeatedly run an ordinary SAT solver as a black box and find satisfying assignments one by one.
A specific procedure is as follows.
\begin{enumerate}
\item Run a SAT solver with a CNF formula $\psi$.\label{enum:basicform1}
\item If $\psi$ is unsatisfiable, halt.
\item Report a found total satisfying assignment $\nu$.
\item Compute the clause $C$ of the form $\set{x}{\nu(x)=0}\cup\set{\lnot x}{\nu(x)=1}$ in a set notation.\label{enum:basicform4}
\item Add $C$ to $\psi$ and go to Step~\ref{enum:basicform1}.
\end{enumerate}

The clause obtained at Step~\ref{enum:basicform4} is called a \emph{blocking clause}~\cite{McMillan}.
Since the blocking clause $C$ is the complement of the term corresponding to $\nu$, the extended CNF formula $\psi\cup\{C\}$ is not satisfied by $\nu$ in a later search, thereby a new solution will be found in each repetition.
Furthermore, since $\nu$ is total, no assignment other than $\nu$ is blocked.

\begin{example}\label{ex:blocking}
Execute the procedure with the CNF formula:
\[\psi=(x_1\lor \lnot x_2)\land (x_2\lor \lnot x_3)\land (x_3\lor \lnot x_1).\]
Suppose a solver returns the satisfying assignment $\lnot x_1,\lnot x_2,\lnot x_3$.
The blocking clause $x_1\lor x_2\lor x_3$ is added to $\psi$:
\[\psi=(x_1\lor \lnot x_2)\land (x_2\lor \lnot x_3)\land (x_3\lor \lnot x_1)\land (x_1\lor x_2\lor x_3).\]
The solver then returns the satisfying assignment $x_1,x_2,x_3$.
The blocking clause $\lnot x_1\lor \lnot x_2\lor \lnot x_3$ is added to $\psi$:
\[\psi=(x_1\lor \lnot x_2)\land (x_2\lor \lnot x_3)\land (x_3\lor \lnot x_1)\land (x_1\lor x_2\lor x_3)\land(\lnot x_1\lor \lnot x_2\lor \lnot x_3).\]
This time a solver returns UNSAT, which means all satisfying assignments are found.
\end{example}

Since a blocking clause has size equal to the number of variables, unit propagation is likely to slow down.
Hence, it is arguably better to consider blocking clauses that consist only of decisions, which we call a \emph{decision-based blocking clause} for convenience.
Since decisions determine the other assignments, decision-based blocking clauses have the same effect as those of all literals.
However, to know which literals are decisions, we need to modify a solver code.
Algorithm~\ref{alg:bc_solver} is a pseudo code obtained by modifying Algorithm~\ref{alg:cdcl}.
For convenience, we simply call this \emph{blocking procedure} to distinguish it from the other procedures presented later.
Only lines 5,12-16 are changed.
At line 5 and 12, a solver halts because $dl\leq 0$ means that all variables are implied without any decision and all solutions are found.
At line 15, all assignments except for those determined without any decision are canceled, and at line 16 a solver backtracks to the root level.

A decision-based blocking clause only blocks a single assignment and there are as many number of blocking clauses as total satisfying assignments.
Since they are stored, it is likely to result in space explosion and slow down of unit propagation.
This is considered as a serious issue since unit propagation in modern SAT solvers occupies the majority of a whole processing time.
Another disadvantage is that whenever a solution is found, a solver is enforced to restart from scrach with an extended CNF formula, not resuming search.

On the other hand, the blocking clause-based implementation might be considered as a good choice for such instances as even one solution is hard to find or for instances with a small number of solutions. 
It is easily implementable, because the blocking clause mechanism can be realized outside a solver or with a small modification on a solver code as demonstrated in Algorithm~\ref{alg:bc_solver}.
We can benefit from powerful techniques of modern SAT solvers such as conflict-driven clause learning, non-chronological backtracking, and so on.

\begin{algorithm}[t]
\SetAlgoNoLine
\KwIn{a CNF formula $\psi$, an empty assignment $\nu$.}
\KwOut{all satisfying assignments.}

$dl              \leftarrow 0$\tcc*[r]{Decision level}
\While{true}{
    $\nu \leftarrow \propagate{\psi}{\nu}$\tcc*[r]{Deduce stage}
    \eIf{conflict happens}
    {
        \lIf{$dl \leq 0$}{halt}
        $bl \leftarrow \analyze{\psi}{\nu}$\tcc*[r]{Diagnose stage}
        $\nu\leftarrow\set{(x,v)\in\nu}{\delta(x) \leq bl}$\;
        $dl       \leftarrow bl$\;
    }
    {
        \eIf{all variables are assigned values}
        {
            report $\nu$\;
            \lIf{$dl \leq 0$}{halt}
            compute a blocking clause $C$ from $\nu$\;
            $\psi \leftarrow \psi\cup\{C\}$\;
            $\nu\leftarrow\set{(x,v)\in\nu}{\delta(x) = 0}$\;
            $dl   \leftarrow 0$\;
        }
        {
            $dl          \leftarrow dl+1$\tcc*[r]{Decide stage}
            select an unassigned variable $x$ and a value $v$\;
            $\nu      \leftarrow \nu\cup\{(x,v)\}$\;
        }
    }
}
\caption{Blocking procedure, where $\delta(x)$ denotes the decision level of $x$.}
\label{alg:bc_solver}
\end{algorithm}

\begin{remark}
Once blocking clauses are added to a CNF formula, they are not deleted afterward and must not be treated in the same way as conflict clauses.
Otherwise, solutions would be rediscovered many times, which is not allowed in the paper.
\end{remark}

\subsubsection{Simplifying Satisfying Assignments}~\label{subsubsec:simplify}
Simplification of satisfying assignments is to obtain from a satisfying assignment $\nu$ to a CNF formula $\psi$ a smaller assignment\footnote{We say that an assignment $\nu'$ is smaller than $\nu$ if all variables $x$ assigned values in $\nu'$ are also assigned values in $\nu$ and their values coincide, i.e., $\nu(x)=\nu'(x)$.} $\nu'$ that still makes $\psi$ evaluate to $1$.
This is done by canceling assignments to \emph{redundant} variables in $\nu$, where redundant means either value is assigned without effect on the value of $\psi$.
A simplified assignment is partial in general and it represents a set of total assignments, including the original assignment and possibly other satisfying assignments.
\emph{Variable lifting} refers to a number of such simplification techniques~\cite{Ravi}.
Since this topic is well-summarized in the literature~\cite{1562927}, we do not go into details.
The interested readers are referred to it, as well as the references therein.
For recent results, see also the literature~\cite{6733111}.

Simplification allows us to obtain from a single solution possibly exponentially many solutions in a compact form, i.e., as a partial assignment.
It is desirable if we can obtain a partial assignment of minimum size, however minimization is known to be computationally hard, and in practice we have to compromise with near-minimum assignment by means of approximation\footnote{Even computing a minimal satisfying assignment requires quadratic time, which is still expensive.}.
To combine simplification with the blocking mechanism of Algorithm~\ref{alg:bc_solver}, it suffices to perform simplification just before line 13 and then to take complement of decisions in a simplified assignment.
It should be noted that if a simplified blocking clause is empty, that is, all variables except for implied ones turn out to be redundant, then it means that all remaining solutions are covered.
Thus, in that case, a solver must be halted.
Thanks to simplification, the number of blocking clauses may be largely reduced, which leads to a good effect on unit propagation at the cost of performing simplification.

\begin{example}\label{ex:simplify}
Consider the CNF formula $\psi$ given in Example~\ref{ex:cnf}.
The assignments $\lnot x_5@1, \lnot x_6@1,x_3@2, x_1@2, x_4@2, x_2@3$ were given in this order.
The last decision is redundant.
By removing it, we obtain the partial assignment $\lnot x_5@1, \lnot x_6@1,x_3@2, x_1@2, x_4@2$.
Consider the decision assignments $\lnot x_5@1$ and $x_3@2$, by which the other variables are implied or can be assigned either value.
By flipping those assignments, we obtain the simplified blocking clause $C_6=x_5\lor \lnot x_3$.
\end{example}

\subsubsection{Continuing Search}~\label{subsubsec:continue}
In Algorithm~\ref{alg:bc_solver}, whenever a solution is found, a solver is enforced to backtrack to the root level.
After that, due to a variable selection heuristic, a different assignment will be examined and a region of the search space in which a solution was just found remains incomplete, which may give rise to unnecessary propagations and conflicts in a later search for that region.
Restart is, however, essential in AllSAT solving in particular when simplification is used.
Indeed after simplification is performed and a new clause is added, the state of implications such as which literals are decisions becomes inconsistent, and it is necessary to deduce implications again.
It is not straightforward to answer how to continue search~\cite{jin}.

The problem of over-canceling due to backtracking was addressed~\cite{DBLP:conf/sat/PipatsrisawatD07}, and a simple technique, called \emph{progress saving}, that stores recent canceled decisions in an array and simulates them after backtracking was proposed.
Specifically, any time a solver enters into the decide stage, it checks if an assignment to a selected variable is stored, and it simulates the previous decision if exists; otherwise, it follows a default heuristic.
Although this technique was proposed in the context of SAT, it is also applicable to AllSAT.

\begin{example}\label{ex:continue}
Continuing Example~\ref{ex:simplify}, suppose that we added the blocking clause $C_6$ to $\psi$ and backtracked to the root level with progress saving enabled.
At this point, all assignments above the root level are canceled, yet the previous decisions $(x_5,0), (x_3,1)$ are stored in an array.
If $x_5$ or $x_3$ is selected, the previous decision is made again. 
\end{example}

\subsubsection{Implementation}~\label{blocking_implemenation}
We implemented 4 programs based on blocking procedure according to whether simplification and continuation techniques are selected or not.

For a simplification technique, we used a method related to set covering model~\cite{1562927} and decision-based minimal satisfying cube~\cite{6733111}.
For the sake of efficiency, near-minimal satisfying assignments are computed.
A basic idea is that given a total assignment, we select as a small number of decision variables as make a CNF formula evaluate to $1$.
This is done in the following way:
\begin{enumerate}
\item select all decision variables that related to implications of at least one variable;\label{enum:naivesimpl1}
\item for each clause that is not satisfied by selected variables, select arbitrary decision variable that makes the current clause satisfied.\label{enum:naivesimpl2}
\end{enumerate}
A simplified satisfying assignment then consists of assignments to the selected decision variables and all implied variables.
By flipping the assignments to the selected decisions, we obtain a blocking clause, which blocks all total assignments represented by the simplified assignment.


We implemented a continuation technique so that before backtracking at line 15, all decisions are stored in an array and after backtracking, we simulate these decisions whenever possible in the order of their decision levels.
Clearly, not all decisions are assumed due to blocking clause, and conflict or contradiction to the previous decision will happen.
It is this point where a solver continues search, and we will then enter into the conflict resolution or the decide stage.

\subsection{Non-blocking Solvers}~\label{subsec:nobcbased}

\subsubsection{Overview}~\label{subsubsec:backtracking}
We give a basic idea for an AllSAT procedure without the aid of blocking clauses.
Like blocking procedure, we modify Algorithm~\ref{alg:cdcl}.
A main feature is to employ chronological backtracking instead of non-chronological backtracking.
The chronological backtracking used here is a bit different from the ordinary one described in Section~\ref{subsec:satsolvers}.
As shown in Fig.~\ref{alg:bt}, only differences are to insert the flipped decision $(x,\bar{v})$ in $\nu$ and register it to an implication graph so that it has no incomming arc.
This is because there is no reason that implied flipped decisions because of the absence of blocking clauses.

We say that a literal, seen as an assignment, \emph{has NULL antecedent} if it has no reason that implied it and there is no incomming arc in an implication graph.
The chronological backtracking given above is hereafter abbreviated as \emph{BT} for convenience, and it is performed by the function $backtrack$.
We collectively call a number of procedures for AllSAT solving based on BT \emph{non-blocking procedure} in contrast to blocking procedure presented in Section~\ref{subsec:bcbased}.
If there is no fear of confusion, chronological backtracking always means BT.

\begin{algorithm}[t]
\SetAlgoNoLine
\KwIn{an assignment $\nu$, the current decision level $dl$.}
\KwOut{updated objects $\nu$, $dl$.}
        $(x,v) \leftarrow$ the decision assignment of level $dl$\;
        $\nu\leftarrow\set{(y,w)\in\nu}{\delta(y) \leq dl-1}$\;
        $dl  \leftarrow dl-1$\;
        $\nu\leftarrow\nu\cup\{(x,\bar{v})\}$, where $\bar{v}$ is the opposite value from $v$\;
        return $\nu$, $dl$\;
\caption{Chronological backtracking in non-blocking procedure, where $\delta(y)$ denotes the decision level of $y$.}
\label{alg:bt}
\end{algorithm}

\begin{figure}[t]
    \small
    \begin{center}
    \begin{minipage}{0.32\hsize}
        \begin{center}
        \begin{tabular}{llll}
        $-1^*$   &        &        &        \\ \hline
        $-20^*$  & $19$   & $9$    & $-15$  \\
        $-8$     & $12$   &        &        \\ \hline
        $-18^*$  & $2$    & $-5$   & $-17$  \\
        $-10$    & $11$   & $-14$  & $13$   \\
        $-6$     & $-3$   & $7$    & $4$    \\
        $16_*$   &        &        &        \\
        \end{tabular} 
        \hspace{1em} (a) Solution Found
        \end{center}
    \end{minipage}
    \begin{minipage}{0.32\hsize}
        \begin{center}
        \begin{tabular}{llll}
        $-1^*$   &        &        &        \\ \hline
        $-20^*$  & $19$   & $9$    & $-15$  \\
        $-8$     & $12$   & $18_*$ &        \\ \hline
        $-16^*$  & $-6$   & $11$   & $2$    \\ 
        $-7$     & $13$   & $-5$   & $-3$   \\
                 &        &        &        \\
                 &        &        &        \\
        \end{tabular} 
        \hspace{1em} (b) Conflict
        \end{center}
    \end{minipage}
    \begin{minipage}{0.32\hsize}
        \begin{center}
        \begin{tabular}{llll}
        $-1^*$   &        &        &        \\ \hline
        $-20^*$  & $19$   & $9$    & $-15$  \\
        $-8$     & $12$   & $18_*$ & $6$    \\
        $-16$    & $-3$   &        &        \\
                 &        &        &        \\
                 &        &        &        \\
                 &        &        &        \\
        \end{tabular} 
        \hspace{1em} (c) Conflict
        \end{center}
    \end{minipage}
    \caption{Snapshots of a non-blocking solver's state in the three different cases (a),(b), and (c).
Assignments are given from left to right, top to bottom, separated by line for each level.
Integers $i$ specify assignments such that the variable $x_{|i|}$ is assigned the value $0$ if $i<0$ and $1$ if $i>0$.
Decision assignments have an asterisk as a superscript.
Non-decision assignments with NULL antecedent have an asterisk as a subscript.
}\label{fig:propagationstack}
    \end{center}
\end{figure}

An important point in this approach is how we make BT compatible with conflict-driven clause learning.
Consider a conflict graph in clause learning phase.
Due to BT, a conflict graph may contain several roots, i.e. assignments with NULL antecedent, in the same decision level.
See for example the literals $\lnot x_{20}$ and $x_{18}$ in Fig.~\ref{fig:conflictgraph3}.
Since an ordinary first UIP scheme commonly assumes a unique root in the same decision level, implementations based on that assumption get stuck in non-decision literals with NULL antecedent.
To resolve this problem, two techniques are presented later.

\begin{example}\label{ex:propagationstack}
Look at Fig.~\ref{fig:propagationstack}.
(a) All variables are assigned values without conflict, which means a solution is found.
(b) Following BT, all assignments of level $3$ are canceled and the flipped decision $x_{18}$ is inserted as a non-decision assignment at level $2$.
Since no propagation takes place, a new decision $\lnot x_{16}$ is made and a subsequent propagation results in conflict.
The decision is the only one that has NULL antecedent in the current decision level, and ordinary first UIP scheme suffices in this case.
(c) From the conflict we have just met, the conflict clause $x_6\lor x_1\lor \lnot x_{18}\lor \lnot x_{12}$ is learnt, and a solver backtracks to level $2$.
\footnote{In this case, there is no oppotunity to rediscover solutions by ordinary chronological backtracking, and hence the flipped decision $x_{16}$ is not inserted. This technique is explained in more detail when non-chronological backtracking with level limit is introduced.}
The assignment $x_6$ is implied by the conflict clause.
A subsequent propagation results in conflict again.
This time, there is a non-decision assignment with NULL antecedent in the same decision level.
\end{example}

\begin{algorithm}[t]
\SetAlgoNoLine
\KwIn{a CNF formula $\psi$, an empty variable assignment $\nu$.}
\KwOut{all satisfying assignments.}

$dl              \leftarrow 0$\tcc*[r]{Decision level}
$lim             \leftarrow 0$\tcc*[r]{Limit level}
\While{true}{
    $\nu \leftarrow \propagate{\psi}{\nu}$\tcc*[r]{Deduce stage}
    \eIf{conflict happens}
    {
        \lIf{$dl \leq 0$}{halt}
        $(\nu,dl,lim) \leftarrow \resolve{\psi}{\nu}{dl}{lim}$\tcc*[r]{Resolve stage}
    }
    {
        \eIf{all variables are assigned values}
        {
            report $\nu$\;
            \lIf{$dl \leq 0$}{halt}
            $(\nu,dl)\leftarrow \backtrack{\nu}{dl}$\;
            $lim \leftarrow dl$\;
        }
        {
            $dl   \leftarrow dl+1$\tcc*[r]{Decide stage}
            select an unassigned variable $x$ and a value $v$\;
            $\nu  \leftarrow \nu\cup\{(x,v)\}$\;
        }
    }
}
\caption{Non-blocking procedure with decision level-based first UIP scheme.}
\label{alg:memeff_solver}
\end{algorithm}

A major advantage of non-blocking approach is that no matter how many solutions exist, a performance of unit propagation does not deteriorate thanks to the absence of blocking clauses.
Instead, it has to find total satisfying assignments one by one.
Hence, there is a limit in the number of solutions to be generated in a realistic mount of time.

\subsubsection{Sublevel-based First UIP Scheme}~\label{subsubsec:sublevel}
Grumberg et al. introduced the notion of \emph{sublevels} and presented a sublevel-based first UIP scheme that is compatible with non-blocking approach~\cite{grumberg}.
A basic idea is to divide a single decision level into sublevels.
Specifically, a new sublevel is defined whenever BT is performed, and sublevels are undefined as their decision levels are undefined.
An ordinary first UIP scheme can then be applied in the current sublevel.

A conflict clause obtained by this approach may contain many literals that are below the current sublevel yet in the same decision level.
Among literals in that conflict clause, those with NULL antecedent are necessary for avoiding rediscovery of solutions and can not be removed if exist, however it is expected that other literals are reduced further.

\subsubsection{Decision Level-based First UIP Scheme*}~\label{subsubsec:decisionlevel}
We present an alternative first UIP scheme that need not require sublevels.
Our scheme can be realized with a small modification: it is simply not to stop at literals with NULL antecedent and attempt to find the first UIP in the current decision level.
A specific procedure is to traverse a conflict graph from $\kappa$ in the reverse order of implications and construct a conflict clause $C$, repeating the following procedure until the first UIP appears:
\begin{enumerate}
\item if the current literal is below the current decision level, add the negated literal to $C$ and do not go up through the incomming arcs of the current literal;
\item if the current literal has NULL antecedent, add the negated literal to $C$;\label{enum:scheme}
\item for the other case, go up through the incoming arcs of the current literal if their source vertices have not yet been visited.\label{enum:scheme}
\end{enumerate}
After the first UIP is found, add the negated literal to $C$.

Compared to the sublevel-based first UIP scheme, the decision level-based scheme might be considered better.
First of all, it is simple.
Secondly, a conflict clause contains unique literal from the current decision level except for those with NULL antecedent.
However, it should be noted that conflict clauses obtained by the decision level-based scheme are not necessarily smaller, because the unique implication point is further from $\kappa$ and thus a conflict clause may contain more literals below the current decision level.
A pseudo code for the modified scheme is omitted, because modification would be straightforward.

\begin{figure*}[t]
\begin{center}
\includegraphics[width=12cm]{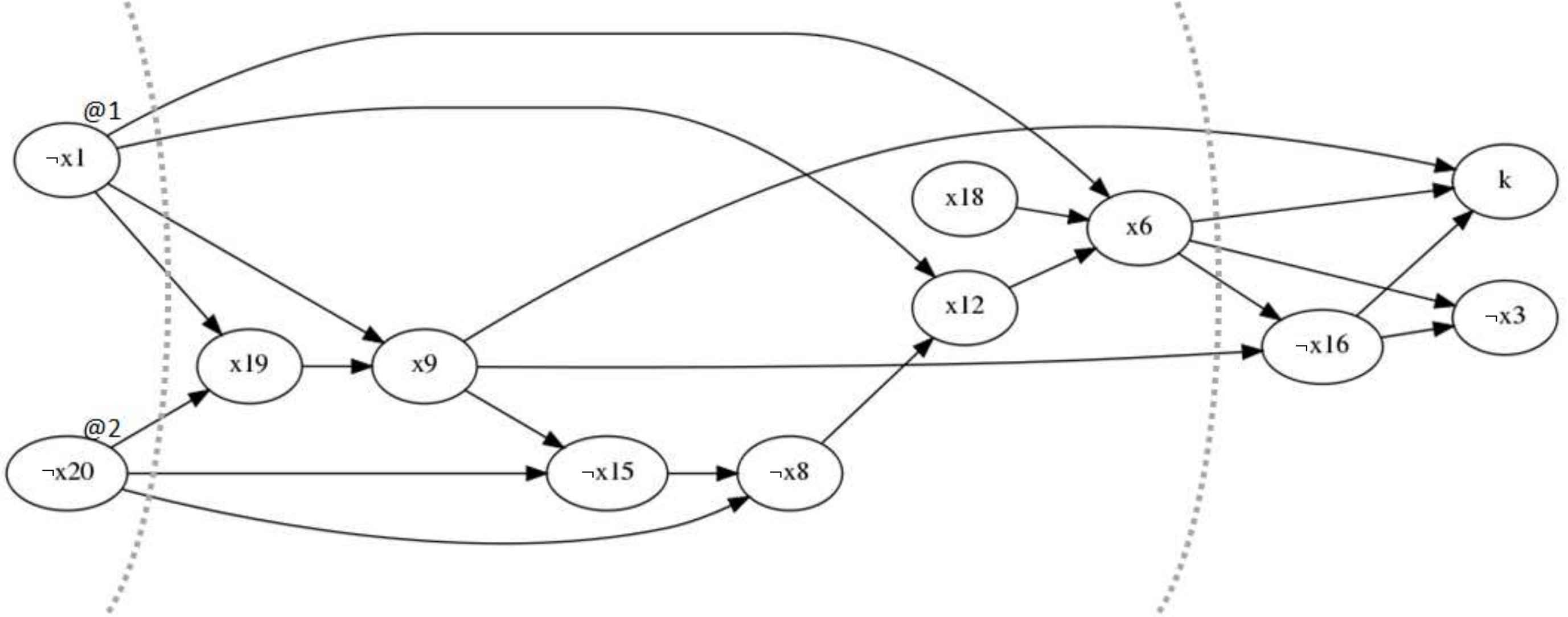}
\end{center}
\caption{An implication graph for the conflict case (c) in Fig.~\ref{fig:propagationstack}.}\label{fig:conflictgraph3}
\end{figure*}

\begin{example}\label{exam:conflictclauses}
Continuing Example~\ref{ex:propagationstack}, consider the conflict case (c) in Fig.~\ref{fig:propagationstack}.
Figure~\ref{fig:conflictgraph3} illustrates the difference of the two schemes: the sublevel-based scheme finds $x_6$ as a first UIP and learns $\lnot x_6\lor \lnot x_9$ as a conflict clause, while the decision level-based scheme finds $\lnot x_{20}$ as a first UIP and learns $x_{20}\lor x_1 \lor \lnot x_{18}$ as a conflict clause.
Note that the conflict clause in either case does not become unit after backtracking, though it does not menace algorithmic correctness because of the flipped decision $x_{20}$.
\end{example}

\begin{remark}
Recall that when the function $analyze$ is performed, an assignment to its unit literal is set to the current assignment function $\nu$.
In non-blocking procedure, a conflict clause is not necessarily unit as seen in Example~\ref{exam:conflictclauses}.
Hence, the function $analyze$ in non-blocking procedure only adds a conflict clause, and it does not consider an induced assignment.
\end{remark}

Algorithm~\ref{alg:memeff_solver} is a pseudo code for the non-blocking approach using clause learning with the decision level-based first UIP scheme.
Recall that the function $backtrack$ is to backtrack chronologically, following Algorithm~\ref{alg:bt}.
There are some choices for it, where a generic function, named $resolve$, is called.
A simple way of realizing the resolve stage is to perform clause learning, based on either first UIP scheme and then perform BT.
More elaborate methods are introduced later, with which $resolve$ can also be replaced.
Since the variable $lim$ is used in one of those methods, it is introduced when needed.

\subsubsection{Conflict Directed Backjumping}~\label{subsubsec:cdb}
Grumberg et al. augmented non-blocking approach with conflict resolution by means of a restricted non-chronological backtracking~\cite{grumberg}.
Their backtracking method can be considered as a form of \emph{conflict directed backjumping} (\emph{CBJ} for short).
CBJ has been studied as one of tree search algorithms for constraint satisfaction problem~\cite{DBLP:journals/ci/Prosser93}~\cite{Chen:2001:CBR:1622394.1622397}~\cite{Dechter2002147}.

A basic idea is described below.
Consider the scenario where conflict happens at decision level $i+1$, and after backtracking to level $i$, conflict happens again.
In this case, obtain a conflict clause $cl_1$ from the former conflict, while from the latter conflict, obtain a conflict clause $cl_2$ so that its UIP is $\lnot p$, where $p$ is the first UIP in the former analysis.
Perform resolution of $cl_1$ and $cl_2$ and obtain a resulting clause $cl_3$.
After that, backtrack to the level preceding the highest level in $cl_3$.

A pseudo code is given in Algorithm~\ref{alg:cbj}, which is almost faithfully rephrased in our setting from the code given in the literature~\cite{grumberg}.
The call of $propagate$ at the end of the \textbf{while} loop was not explicitly written in the original code, which we consider necessary.
At this call, unit propagation considers the effect of the recent flipped decision inserted as a result of $backtrack$.
Note that the effect of an assignment implied by the recent conflict clause is not considered here, because our non-blocking procedure assumes that the function $analysis$ simply records a conflict clause and does not insert an implied assignment, which is separately inserted at line 12.
The halt means that non-blocking procedure is halted too.

\begin{algorithm}[t]
\SetAlgoNoLine
\KwIn{a CNF formula $\psi$, an assignment $\nu$, the current decision level $dl$.}
\KwOut{updated objects $\nu$, $dl$.}
$stack\leftarrow$ an empty stack\;
\While{true}
{
    \uIf{conflict happens}
    {
        \lIf{$dl\leq 0$}{halt}
        $\analyze{\psi}{\nu}$\;
        push the learnt conflict clause into $stack$\;
        $(\nu,dl)\leftarrow\backtrack{\nu}{dl}$\;
    }
    \uElseIf{$stack$ is not empty}{
        $cl_1\leftarrow$ the clause popped from $stack$\;    
        \If{$cl_1$ is a unit clause}
        {
            $unit\leftarrow$ the unit literal in $cl_1$\; 
            add $unit$, seen as an assignment, to $\nu$ so that it has antecedent $cl_1$\;
            $\nu\leftarrow\propagate{\psi}{\nu}$\;
            \If{conflict happens}
            {
                \lIf{$dl\leq 0$}{halt}
                $cl_2\leftarrow$ the conflict clause from the recent conflict with $unit$ as UIP\;
                $cl_3\leftarrow$ the resolution of $cl_1$ and $cl_2$\;
                push $cl_3$ into $stack$\;
                $bl\leftarrow$ the highest level in $cl_3$\;
                $(\nu,dl)\leftarrow\backtrack{\nu}{bl}$\;
            }
        } 
    }
    \Else{
        break\;
    }
    $\nu\leftarrow\propagate{\psi}{\nu}$\;
}
return $\nu$, $dl$\;
\caption{Conflict resolution based on conflict-directed backjumping}
\label{alg:cbj}
\end{algorithm}

\subsubsection{Non-chronological backtracking with Level Limit And Its Combination with CBJ*}\label{subsubsec:limitlevel}
We present an alternative conflict resolution by means of non-chronological backtracking with backtrack level limit.
To our knowledge, this method was first presented by~\cite{claspenum}, though it was in the context of answer set programming.
We thus would like to import their idea to our non-blocking procedure.

Since non-blocking procedure does not record blocking clauses, it must not backtrack to arbitrary level, even though a backtracking level is one that is legitimately derived from a conflict clause.
However, we can do this if there is no opportunity of rediscovering found solutions from the derived level.
We use the variable $lim$ that holds a "safe" level to be backtracked, that is, the first level at which the current assignment and the previous satisfying assignment differ.

A pseudo code is given in Algorithm~\ref{alg:limitedbj}.
We call this approach \emph{non-chronological backtracking with level limit}, denoted by \underline{BJ}, where the underline means a backtrack level is limited.
Since $lim$ is always less than or equal to $dl$, BT is performed if and only if $lim=dl$.
If $lim < dl$, backtracking does not entail inserting a flipped decision (see Example~\ref{ex:propagationstack}).
This is because $lim < dl$ implies no solution yet found, and hence no opportunity to rediscover solutions by backtracking.

\begin{algorithm}[t]
\SetAlgoNoLine
\KwIn{a CNF formula $\psi$, an assignment $\nu$, the current decision level $dl$, a level limit $lim$.}
\KwOut{updated objects $\nu$, $dl$, $lim$.}
        $bl \leftarrow \analyze{\psi}{\nu}$\tcc*[r]{Diagnose stage}
        \eIf{$lim < dl$}
        {
            \lIf{$bl < lim$}{$bl=lim$}
            $\nu\leftarrow\set{(x,v)\in\nu}{\delta(x) \leq bl}$\;
            $dl \leftarrow bl$\;
        }
        {
            $(\nu,dl)\leftarrow\backtrack{\nu}{dl}$\;
            $lim \leftarrow dl$\;
        }
        return $\nu$, $dl$, $lim$\;
\caption{Conflict resolution based on non-chronological backtracking with level limit}
\label{alg:limitedbj}
\end{algorithm}

We furthermore present a combination\footnote{It is mentioned~\cite{claspenum} that a combination of conflict-directed backjumping and non-chronological backtracking is proposed. However, their pseudo code almost corresponds to Algorithm~\ref{alg:limitedbj} and it is different from that stated in this paper.} of \underline{BJ} and CBJ.
This is obtained by replacing the \textbf{else} part in \underline{BJ}, in which BT is performed, with CBJ: that is,
\begin{enumerate}
\item if $lim$ is less than the current decision level, then perform \underline{BJ};\label{enum:comb1}
\item otherwise, perform CBJ.\label{enum:comb2}
\end{enumerate}

Step~\ref{enum:comb2} is selected only when $lim$ equals the current decision level: in other words, when conflict happens just after the current assignment diverged from the previous satisfying assignment.
Hence, \underline{BJ} is preferentially applied when one or more  decisions from the diverging point are made.
This is designed with expectation that more decisions are made from the diverging point, more effectively \underline{BJ} prunes search space.
Since \underline{BJ} is likely to be more frequently applied, this approach is denoted by \emph{\underline{BJ}+CBJ}.

\subsubsection{Implementation}~\label{nonblocking_implemenation}
We implemented 8 programs based on non-blocking procedure according to which of the two first UIP schemes is selected and which of the conflict resolution methods, i.e.\ BT, \underline{BJ}, CBJ, or \underline{BJ}+CBJ, is selected, where BT means performing BT after clause learning. 

\subsection{Formula-BDD Caching Solvers}~\label{subsec:bddbased}

\subsubsection{Overview}~\label{subsubsec:formulacaching}
\emph{Formula caching} refers to a number of techniques to memorize formulae to avoid recomputation of subproblems~\cite{Beame:2010:FCD:1714450.1714452}.
Examples include a caching technique in probabilistic planning~\cite{DBLP:conf/aaai/MajercikL98}, conflict clauses in SAT~\cite{Bayardo:1997:UCL:1867406.1867438}~\cite{769433}, component caching and other cachings in \#SAT~\cite{Bayardo00countingmodels}~\cite{1238208}, and blocking clauses in AllSAT~\cite{McMillan}.

Another type of formula caching in which formulae are associated with propositional languages such as FBDD, OBDD, and a subset of d-DNNF has been studied in the context of knowledge compilation~\cite{DBLP:journals/jair/HuangD07}.
Their work revealed a correspondence between exhaustive DPLL search and propositional languages.
They also proposed speeding up compilation by exploiting techniques of modern SAT solvers through the correspondence.
Although exhaustive DPLL search is simply used for efficiency in their compilation approach, compilation in turn can contribute to speeding up exhaustive DPLL search.
Actually, if a CNF formula is compiled into a BDD, all satisfying assignments can be generated simply by traversing all possible paths from the root to the sink node $\top$.
This seems like taking a long way around to AllSAT solving, however thanks to the caching mechanism, recomputation of many subproblems can be saved.
A connection to AllSAT was mentioned, however their primary concern is on the compilation to suitable languages for required queries, not restricted to AllSAT.
To our knowledge, comparisons have been conducted only between various compilers.
An application to an AllSAT solver itself was more explicitly mentioned in the literature~\cite{Toda:2015:BCS:2695664.2695941} and a compiler-based AllSAT solver is released.
However, comparisons with other AllSAT solvers have not been conducted yet and its power remains unknown.
Similar caching techniques appear in other areas such as preimage computation in unbounded model checking~\cite{Sheng:2003:EPC:789083.1022826}~\cite{DBLP:conf/date/LiHS04}~\cite{1386372}, satisfiability~\cite{978-3-540-43977-6}, and discrete optimization~\cite{anderson:mdd}~\cite{discreteopt}.

The paper only deals with the caching method that records pairs of formulae and OBDDs, which we call \emph{formula-BDD caching}.
A formula-BDD caching can be embedded in either blocking procedure or non-blocking procedure\footnote{Only the combination with blocking procedure has been presented in the past work~\cite{darwiche}~\cite{Toda:2015:BCS:2695664.2695941}.}.
This is done without almost any loss of optimizations employed in an underlying procedure.
An exception is that variables must be selected in a fixed order at the decide stage.
This effect is far from negligible in terms of efficiency, as is well-recognized in a single solution SAT.
It is, however, confirmed in experiments that formula-BDD caching solvers exhibit a quite good performance on the whole, and it provides an efficient solution method for instances that have a huge number of solutions and can not possibly be solved by other means.

\subsubsection{Caching Mechanism}\label{subsubsec:formula-bddmechanism}
We give a basic idea of formula-BDD caching, using a simple BDD construction method with formula-BDD caching.
This method is elaborated later by implementing on top of a SAT solver.
We first introduce terminology.
A \emph{subinstance} of $\psi$ in an assignment $\nu$ is the CNF formula derived from $\psi$ by applying all assignments defined in $\nu$ to $\psi$.
The \emph{current subinstance} refers to the subinstance induced by the current assignment.

Consider the following procedure with a CNF formula $\psi$ and an empty variable assignment $\nu$ as initial arguments:
\begin{enumerate}
\item if an unsatisfied clause exists in the current subinstance, then return $\bot$;
\item if all variables are assigned values, then return $\top$;
\item $i\leftarrow$ the smallest index of an unassigned variable;
\item $f_0\leftarrow$ the result obtained by a recursive call with $\psi$ and $\nu\cup\{(x_i,0)\}$;
\item $f_1\leftarrow$ the result obtained by a recursive call with $\psi$ and $\nu\cup\{(x_i,1)\}$;
\item return a node with the label $i$, the references to LO child $f_0$ and to HI child $f_1$;
\end{enumerate}

Since different assignments can yield subinstances that are logically equivalent, we want to speed up the procedure by applying dynamic programming.
To do this, we need to quickly decide if the current subinstance is solved.
If it is unsolved, we compute a BDD for all solutions of the instance and memorize it, associating with the instance.
Otherwise, the result is obtained in a form of BDD and recomputation is avoided.

However, this approach involves the equivalence test of CNF formulae, which is computationally intractable, as it includes satisfiability testing.
Hence, we consider a weaker equivalence test.
That is, we encode subinstances into formulae so that \emph{if two subinstances are not logically equivalent, then the encoded formulae are not identical}.
To decide if the current subinstance is solved, it suffices to search the encoded formula in the set of registered formula-BDD pairs.
All requirements for formula-BDD caching to work is simply the sentence in italic above, and any encoding that meets it will do.
It should be noted that our test is sound in that acceptance always is a correct decision, however if we prioritize efficiency of encoding excessively, logically equivalent subinstances are very likely to result in non-identical formulae, i.e., a wrong decision.

Examples of formula-BDD cachings include those induced by \emph{cutsets} and \emph{separators}~\cite{darwiche}, defined below, and a variant of cutsets~\cite{Toda:2015:BCS:2695664.2695941}.

\begin{definition}
The $i$-th \emph{cutset} of a CNF formula $\psi$ is the set of clauses $C$ in $\psi$ such that $C$ has literals with their underlying variables $x_j$ and $x_k$ satisfying $j\leq i < k$.
The \emph{cutwidth} of $\psi$ is the maximum size of a cutset of $\psi$.
\end{definition}

\begin{definition}
The $i$-th \emph{separator} of a CNF formula $\psi$ is the set of variables $x_j$ such that some clause in the $i$-th cutset has a literal with its underlying variable $x_j$ satisfying $j\leq i$.
The \emph{pathwidth} of $\psi$ is the maximum size of a separtor of $\psi$.
\end{definition}

\begin{figure*}[t]
    \begin{center}
    \begin{minipage}{0.4\hsize}
        \begin{center}
        \includegraphics[width=5.5cm]{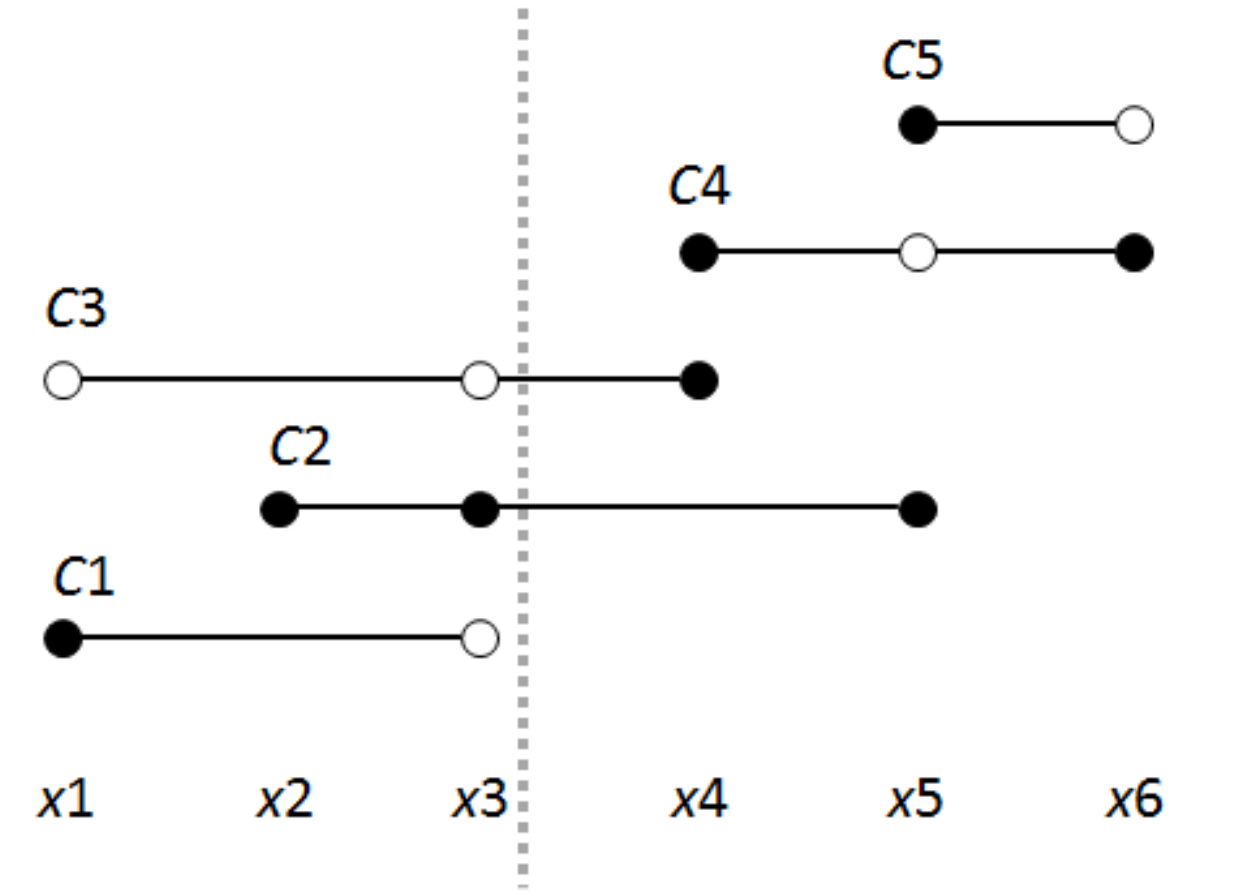}
        \end{center}
    \end{minipage}
    \begin{minipage}{0.55\hsize}
        \begin{center}
        \includegraphics[width=6.5cm]{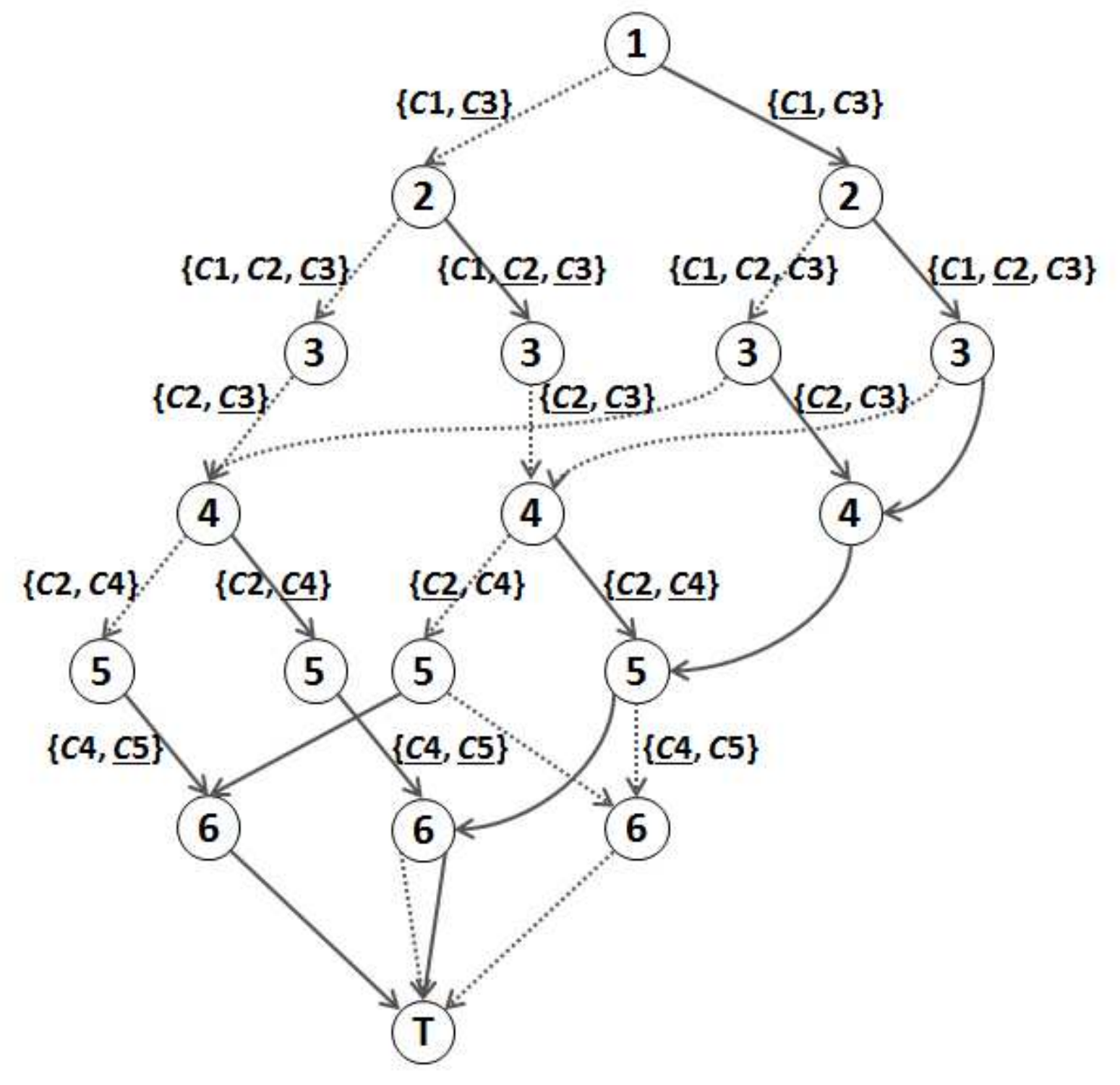}
        \end{center}
    \end{minipage}
    \end{center}
\caption{A CNF formula (left )and an OBDD for the CNF (right), where arcs to the sink node $\bot$ is omitted.
Cutsets are associated with arcs, where underlined clauses mean they are satisfied.}\label{fig:cnfgraph}
\end{figure*}

\begin{example}
Look at the CNF formula $\psi$ illustrated in Fig.~\ref{fig:cnfgraph}.
The $3$-rd cutset of $\psi$ consists of $C_2$ and $C_3$, while the $3$-rd separator of $\psi$ consists of $x_1$, $x_2$, and $x_3$.
The cutwidth and the pathwidth of $\psi$ are both $3$.
\end{example}

The following proposition states that clauses (and variables) in cutsets (and separators) meet the requirement of formula-BDD caching, respectively.
Proof is omitted.
See~\cite{darwiche}.

\begin{proposition}\label{prop:cache}
Let $\psi$ be a CNF formula, where variables are ordered according to their indices.
Let $\nu$ and $\mu$ be assignments with the $i$-th or less variables assigned values and other variables unassigned.
\begin{enumerate}
\item If satisfied clauses in the $i$-th cutset of $\psi$ in $\nu$ coincide with those in $\mu$, then the subinstance in $\nu$ is logically equivalent to that in $\mu$.
\item If variables assigned the value $1$ in the $i$-th separator of $\psi$ in $\nu$ coincide with those in $\mu$, then the subinstance in $\nu$ is logically equivalent to that in $\mu$.
\end{enumerate}
\end{proposition}

\begin{example}
Figure~\ref{fig:cnfgraph} illustrates an OBDD constructed using cutsets as formula-BDD caching.
Cutsets are associated with arcs, and satisfied clauses in them are underlined.
If two arcs have the same set of satisfied clauses, Proposition~\ref{prop:cache} implies that their target vertices can be merged safely.
\end{example}

It should be noted that our weaker equivalence test may reject logically equivalent subintances; the subgraphs in a constructed BDD that correspond to those subinstances are not merged.
This means that a constructed BDD is not fully reduced, i.e., an OBDD, though the OBDD give in Fig.~\ref{fig:cnfgraph} happens to be fully reduced.

An important point of the formula-BDD caching approach is how to balance quality with efficiency in our weaker equivalence test.
The quality here refers to how many correct decisions are made.
Theoretically, it holds that  a correct decision of the separator approach always implies a correct decision of the cutset approach.
On the other hand, the efficiency refers to how much time is taken to create formulae from subinstances, which substantially amounts to evaluating clauses and variables in cutsets and separators, respectively.
In terms of efficiency, evaluating clauses in a cutset would require time linear to the total size of clauses due to lazy evaluation mechanism if it is implemented on top of modern SAT solvers.
On the other hand, evaluating variables in a separator requires time linear to the number of variables.

From the argument above, we can say that for instances with small cutwidth, evaluation cost of cutset is negligible compared to separator and hence cutset is a better choice, and for instances with many clauses, separator should be used instead.

\subsubsection{Embedding Formula-BDD Caching in AllSAT Procedure*}
We demonstrate how to embed formula-BDD caching in a concrete AllSAT procedure.
We take non-blocking procedure as an example.
To our knowledge, this is the first time that the combination is presented.
The other combination, i.e.\ blocking procedure with formula-BDD caching, is omitted.

We have assumed so far that conflict clauses (and blocking clauses in blocking procedure) are added to a CNF $\psi$, but from now on we will assume that they are separately maintained from $\psi$.
This makes $\psi$ unchanged throughout the execution of our AllSAT procedure.
Accordingly, the cutset and the separator in each level are unchanged too.

Algorithm~\ref{alg:bdd_solver} is a pseudo code of non-blocking procedure with formula-BDD caching embedded.
Formula-BDD caching mechanism consists of the encode stage, the extend stage, and the enroll stage.

At the encode stage, the function $makeformula$ receives a CNF $\psi$, the current assignment $\nu$, and the index $i-1$ of the largest assigned variable.
It computes a formula for the current subinstance.
No specific encoding is presented here. See Section~\ref{subsubsec:formula-bddmechanism} for examples of encoding.
Note that if all variables are assigned values, then $i$ must be $\infty$, and hence we have $i-1=\infty$.
In this case, let the function $makeformula$ return $1$, which is the special formula representing true.

At line 15, we search an entry with the key $(i-1,\phi)$ in $S$, where $S$ holds registered formula-BDD pairs.
If it exists, the current subinstance is already solved, and the result is the OBDD $g$ associated with the key, which appears as a subgraph of  $f$.
Hence, at the extend stage, the function $extendobdd$ augments an OBDD $f$ by adding a path from the root of $f$ to the root of $g$, following the current assignment $\nu$.
It returns the pair of the extended OBDD and the added path.
Since this stage is straightforward, we omit a pseudo code of it.

At the enroll stage, we associate formulae for solved subinstances with the corresponding OBDDs and insert these formula-BDD pairs to $S$.
To do this, important points are how to identify when subinstances are solved and how to find their OBDDs.
Let $I$ be a subinstance with the smallest unassigned variable $x_i$.
Since unit propagation is performed at the beginning of each repetition, without loss of generality we assume that $x_i$ is a decision variable.
This means that a formula $\zeta$ has been made from $I$, thereby $(i-1,\zeta)$ is in $T$.
Let $\nu_I$ and $dl_I$ be the assignment and the decision level at this point.
Without loss of generality we assume that there exist one or more solutions extending $\nu_I$, because otherwise, the OBDD for $I$ is not created.
After adding at least one path to $f$, we have the node that corresponds to the decision at variable $x_i$.
Clearly, this node is reachable from the root of $f$ through the path following $\nu_I$, and it is the root of the OBDD for $I$.
When all solutions for $I$ are found, to go out of the exhausted search space, backtracking to a lower level than $dl_I$ is performed, which is directly triggered by an occurrence of conflict implying no solution left or by the discovery of the last solution.
Backtracking to a lower level can also be performed without exhausting all solutions, however this case only happens when no solution of $I$ is yet found.
Hence, we can distinguish them.
Summarizing the above, \emph{if a backtracking level is less than $dl_I$ and at least one solution for $I$ is found, then $I$ is solved, and the root of the OBDD for $I$ is located at the end of the path following $\nu_I$, which is a part of the assignment just before backtracking}.

The function $associate$ is in charge of the enroll stage.
With the observation above, it is called whenever backtracking is performed.
A procedure for it is given as follows.
We scan all nodes $h$ in the most recently added path $\pi$ until the assignment at $h$ taken in $\pi$ contradicts to the current assignment.
Note that scanned nodes $h$ correspond to subinstances $I_h$ such that $I_h$ turns out to be satisfiable and the assignment that induced $I_h$ is a part of the current assignment.
For each scanned node $h$, we test if a backtracking level is less than the decision level of $h$ and a formula $\zeta$ was made\footnote{
This is equivalent to finding an entry $(j-1,\zeta)$ in $T$, where $j$ is the label of $h$.} from $I_h$.
If the test is passed, the pair of $(j-1,\zeta)$ and $h$ is inserted into $S$, where $j$ is the label of $h$.

The function $resolveplus$ behaves in the same way as $reslove$ except that $S$ and $T$ are updated each time backtracking is performed, as in lines 19-20.
 
\begin{figure*}[t]
\begin{center}
\includegraphics[height=4cm]{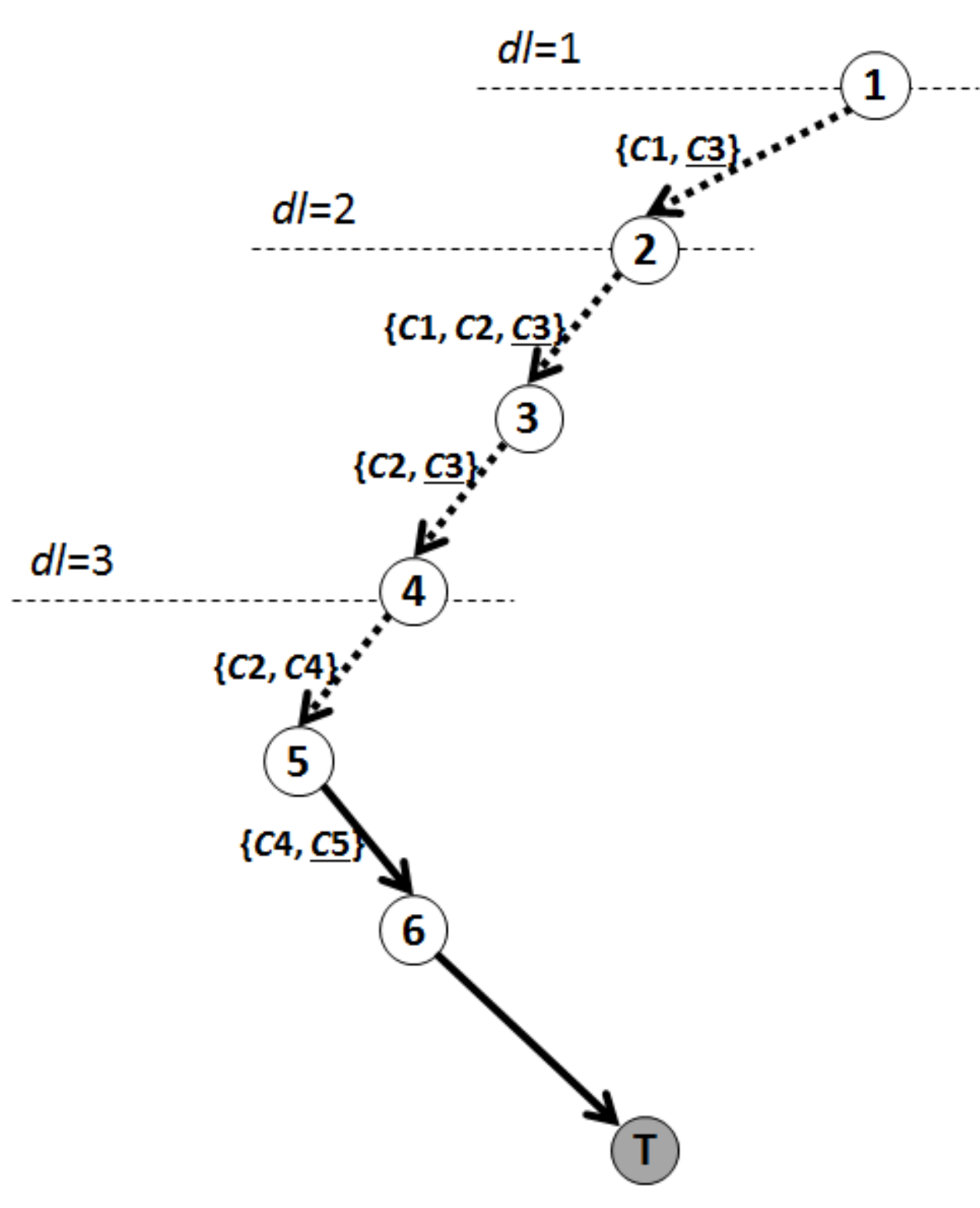}
\includegraphics[height=4cm]{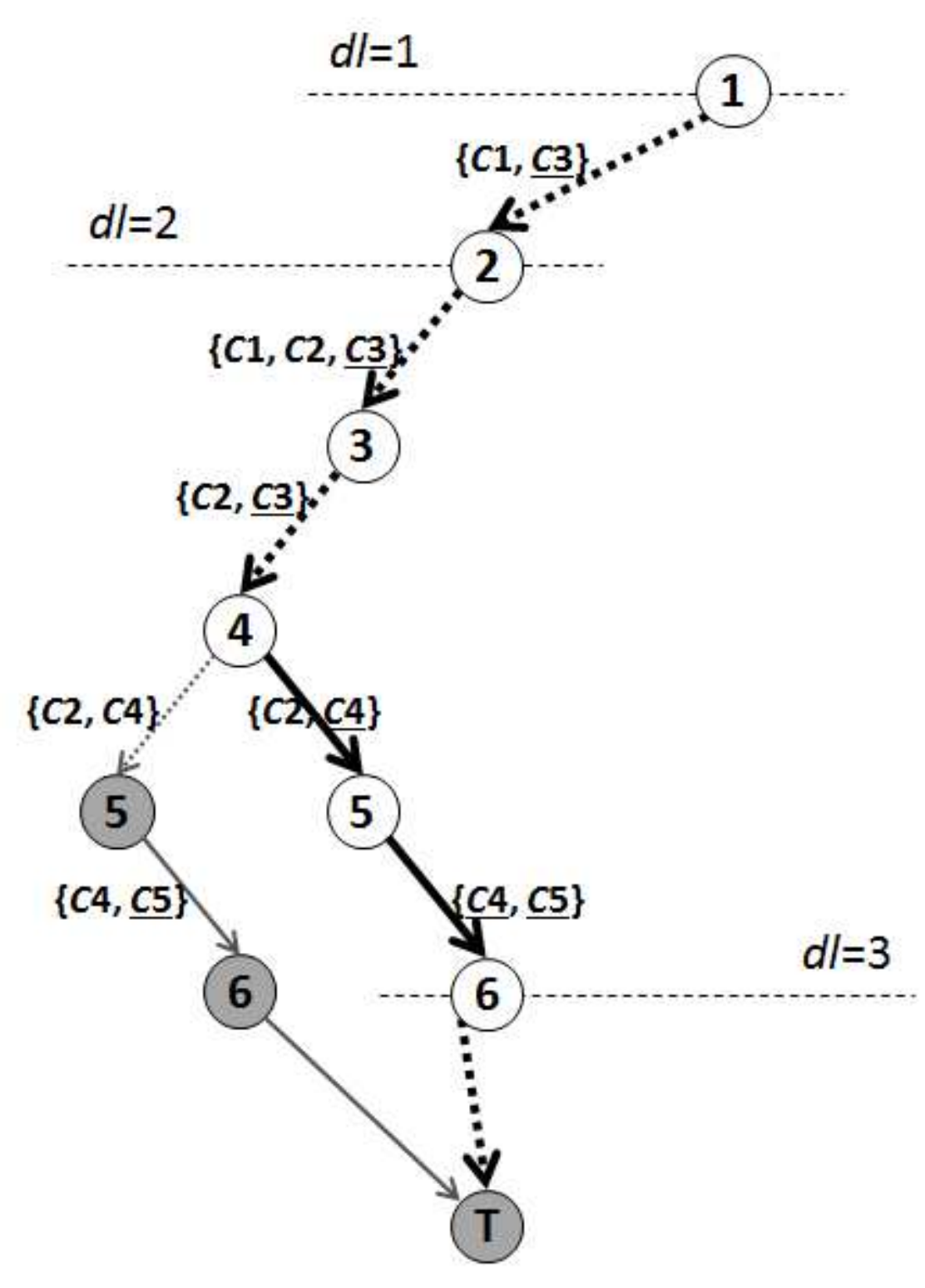}
\includegraphics[height=4cm]{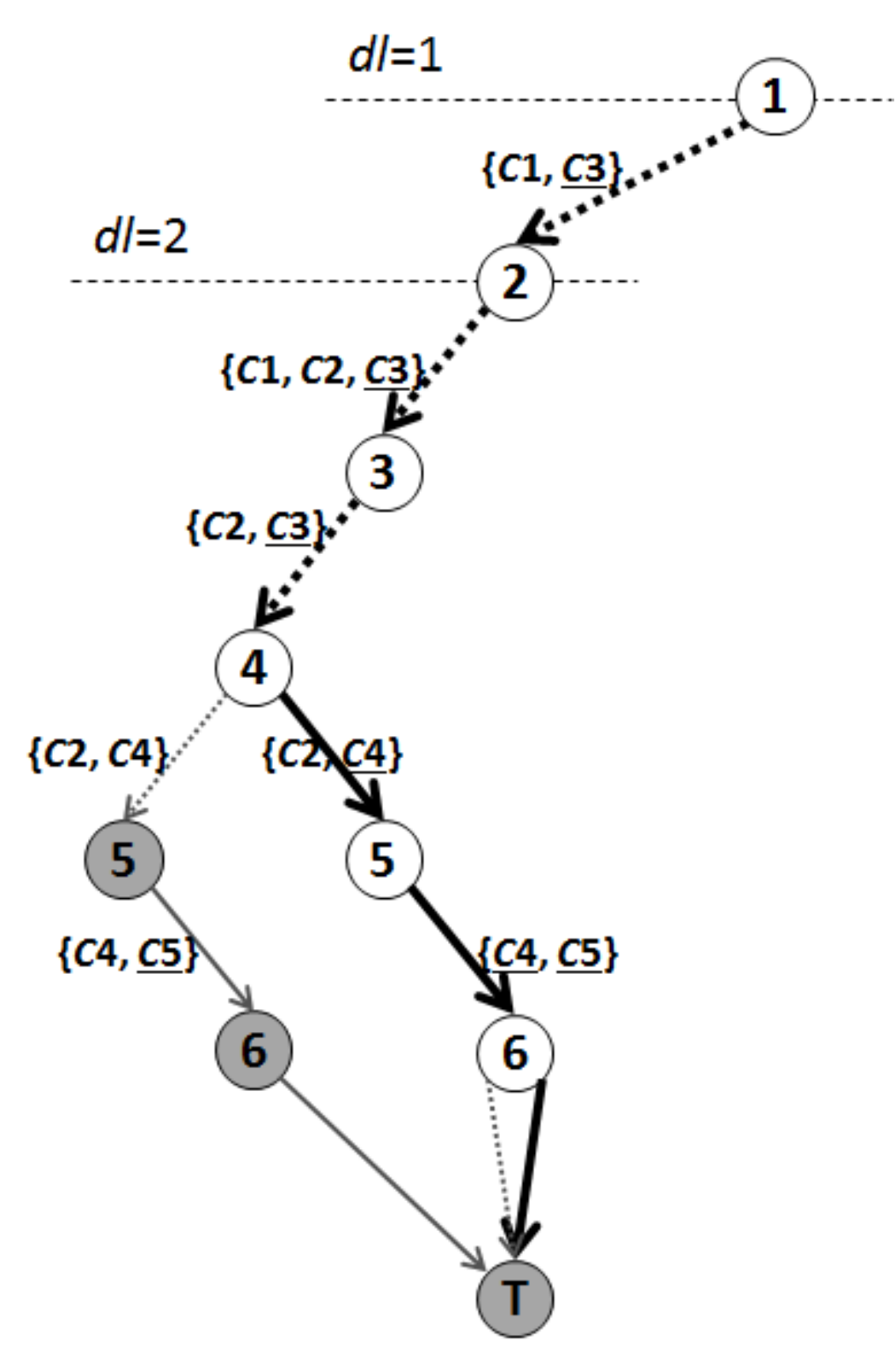}
\includegraphics[height=4cm]{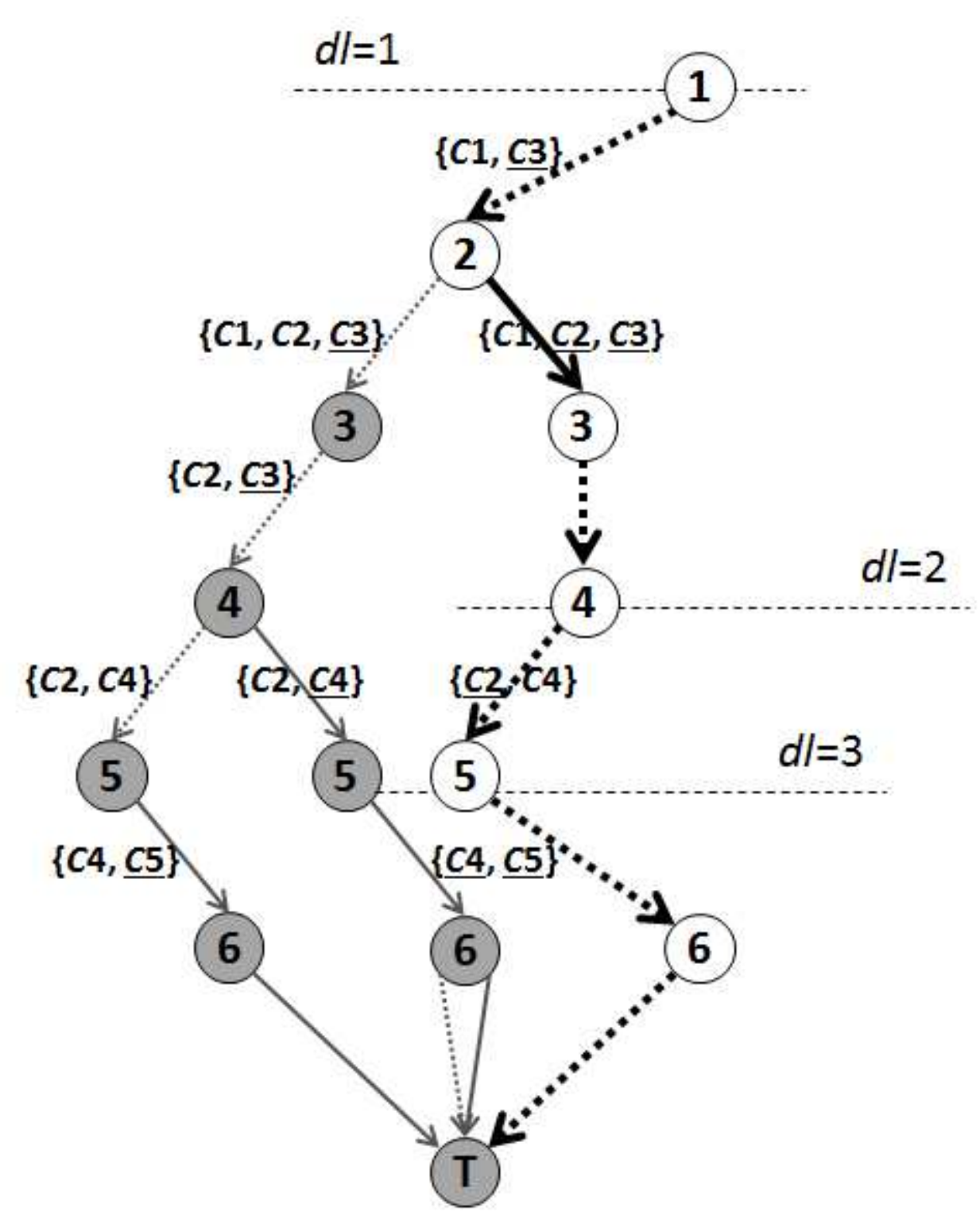}
\end{center}
\caption{
Progress of OBDD construction.
As paths are added one by one, an OBDD is augmented from left to right.
In each step, thick arcs represent a path that is about to be added, and gray nodes mean that the subintances corresponding to them were solved.
}\label{fig:const}
\end{figure*}

\begin{example}
Figure~\ref{fig:const} illustrates how Algorithm~\ref{alg:bdd_solver} constructs an OBDD for the CNF formula $\psi$ given in Fig.~\ref{fig:cnfgraph}.

\end{example}

\begin{algorithm}[t]
\SetAlgoNoLine
\KwIn{a CNF formula $\psi$, an empty variable assignment $\nu$.}
\KwOut{an OBDD $f$ for all satisfying assignments.}

$dl     \leftarrow 0$\tcc*[r]{Decision level}
$lim             \leftarrow 0$\tcc*[r]{Limit level}
$f      \leftarrow \bot$\tcc*[r]{OBDD}
$S      \leftarrow \{((\infty,1), \top)\}$\tcc*[r]{Set of formula-BDD pairs}
$T      \leftarrow \emptyset$\tcc*[r]{Set of formulae with indices}
$\pi    \leftarrow \epsilon$\tcc*[r]{Sequence of BDD nodes}
\While{true}{
    $\nu \leftarrow \propagate{\psi}{\nu}$\tcc*[r]{Deduce stage}
    \eIf{conflict happens}
    {
        \lIf{$dl \leq 0$}{return $f$}
        $(\nu,dl,lim,S,T) \leftarrow \resolveplus{\psi}{\nu}{dl}{lim}{S}{T}$\tcc*[r]{Resolve and enroll stage}
    }
    {
        $i \leftarrow \min\set{j}{\text{$x_j$ is not assigned value}}$\;
        $\phi \leftarrow \makeformula{\psi}{\nu}{i-1}$\tcc*[r]{Encode stage}
        \eIf{an entry with key $(i-1,\phi)$ exists in $S$}
        {
            $g\leftarrow$ the OBDD node associated with $(i-1,\phi)$ in $S$\;
            $(f,\pi) \leftarrow \extendobdd{f}{g}{\nu}$\tcc*[r]{Extend stage}
            \lIf{$dl\leq 0$}{return $f$}
            $S \leftarrow \associate{\pi}{\nu}{S}{T}{dl-1}$\tcc*[r]{Enroll stage}
            $T \leftarrow\set{(j-1,\zeta)\in T}{\delta(x_j) \leq dl-1}$\;
            $(\nu,dl)\leftarrow\backtrack{\nu}{dl}$\;
            $lim \leftarrow dl$\;
        }
        {
            $T\leftarrow T\cup\{(i-1,\phi)\}$\;
            $dl   \leftarrow dl+1$\tcc*[r]{Decide stage}
            select a value $v$\;
            $\nu  \leftarrow \nu\cup\{(x_i,v)\}$\;
        }
    }
}
\caption{Non-blocking procedure with formula-BDD caching, where $\delta(x)$ denotes the decision level of a variable $x$.}
\label{alg:bdd_solver}
\end{algorithm}

\subsubsection{Refreshing OBDDs*}
A constructed OBDD may become too large to be stored on memory, though it would be in practice much better in size than the list representation of all solutions and in the worst case equal to it except for constant factor.
We present a simple technique to resolve this problem.
Let $n$ be the number of variables.
We introduce a threshold $\theta$ of an OBDD size, where $\theta > n$. 
Insert the following procedure after an BDD is extended and backtracking is performed.
If the size of an OBDD is larger than or equal to $\theta-n$, then the current OBDD $f$ is dumped to a file in a secondary storage, and all objects $f$, $S$, $T$, and $\pi$ of formula-BDD caching mechanism are refreshed with initial states.
Since formula-BDD caching is almost independent of the underlying non-blocking procedure, after refreshing, the procedure simply attempts to examine unprocessed assignments with formula-BDD caching empty.

\subsubsection{Implementation}~\label{bdd_implemenation}
We implemented 2 programs on each non-blocking procedure and 2 programs on each blocking procedure according to which formula-BDD caching is selected, cutset or separator.

\section{Experiments}\label{sec:exp}

\subsection{Implementation And Environment.}
All solvers are implemented in C on top of MiniSat-C v1.14.1~\cite{978-3-540-20851-8}.
Clasp 3.1.2 was taken from Potassco project\footnote{Potassco, the Potsdam Answer Set Solving Collection, bundles tools for Answer Set Programming developed at the University of Potsdam,\url{http://potassco.sourceforge.net/}, accessed on 13rd Sept., 2015.}.
As far as we are aware, clasp~\cite{clasp}, picosat~\cite{DBLP:journals/jsat/Biere08}, and relsat~\cite{Bayardo:1997:UCL:1867406.1867438} are only SAT solvers which support the enumeration of all solutions.
Among them, we used clasp for the comparison, because it achieved better performance than picosat, and relsat does not support quiet mode in solution generation and generated solutions may be too large to be stored. 

All experiments were performed on 2.13GHz Xeon\textregistered E7- 2830 with 512GB RAM, running Red Hat Enterprise Linux 6.3 with gcc compiler version 4.4.7.
In the execution of each AllSAT solver, time limit and memory limit were set to 600 seconds and 50GB, respectively.
If either limit is exceeded, the solver is enforced to halt.
All solvers simply touch found solutions and never output them.

The types of compared solvers are a blocking solver, a non-blocking solver, a formula-BDD caching solver, and clasp.
The first three types have some variations according to which techniques are used (see the end of each subsection in Section~\ref{sec:tech}).
Among solvers of the same type, we selected a solver with the most solved instances.
The selected solvers, called \emph{representative solvers}, are as follows.
\begin{itemize}
\item \textbf{Blocking\_NoSimple\_Cont}: the blocking solver with simplification unselected and continuation selected.
\item \textbf{NonBlocking\_DLevel\_BJ}: the non-blocking solver with decision level first UIP scheme and non-chronological backtracking with level limit both selected.
\item \textbf{BDD\_Cut\_NonBlocking\_DLevel\_BJ}: the formula-BDD caching solver with cutset caching selected and it is implemented on top of NonBlocking\_DLevel\_BJ.
\end{itemize}
Throughout the section, if there is no fear of confusion, they are abbreviated as \textbf{Blocking}, \textbf{NonBLocking}, and \textbf{BDD}.
Notation for solvers with other configurations is introduced in the same way.

It is known that variable orderings significantly affect the performance of BDD compilation.
Hence, we used the software MINCE version 1.0~\cite{DBLP:journals/jucs/AloulMS04} to decide a static variable order before the execution of formula-BDD caching solvers.
The execution of MINCE failed for some instances, and for that case, we used the original order.
The time required for deciding a variable order is included.
Although for some instances time limit exceeded in preprocessing, it was negligible for many instances.

\subsection{Problem Instances.} 
We used total 2867 CNF instances (all satisfiable), which are classified as follows.
\begin{itemize}
\item \textbf{satlib}: SATLIB benchmark problems (2707 instances), taken from SATLIB website\footnote{{SATLIB} - The Satisfiability Library, by Holger H Hoos and Thomas St\"{u}tzle at Darmstadt University of Technology, \url{http://www.cs.ubc.ca/~hoos/SATLIB/benchm.html}, accessed on 16th May, 2014.}.
\item \textbf{sc14}: SAT Competition 2014 benchmarks, application track (56 instances) and crafted track (65 instances), taken from SAT Competition 2014 website\footnote{SAT Competition 2014, \url{http://www.satcompetition.org/2014/description.shtml}, accessed on 8th Sept., 2015.}.
\item \textbf{iscas}: ISCAS85 and 89 circuit benchmarks in DIMACS CNF format (39 instances), taken from TG-Pro website\footnote{TG-Pro - A SAT-based ATPG System, by Huan Chen and Joao Marques-Silva, \url{http://logos.ucd.ie/web/lib/exe/fetch.php?media=benchmarks:circuit-cnf-tgpro-20110317.tgz}, accessed on 16th May, 2014}.
\end{itemize}
Among instances released in each repository, we selected all instances such that satisfiability could be decided in 600 seconds by either one of the SAT solvers clasp 3.1.2, glucose4, minisat 2.2, and minisat 1.3, and its result was satisfiable.
For \textbf{satlib} and \textbf{sc14}, random instances are excluded.

\subsection{Comparison of Running Time.}
\begin{figure*}[t]
\begin{center}
\includegraphics[width=10cm]{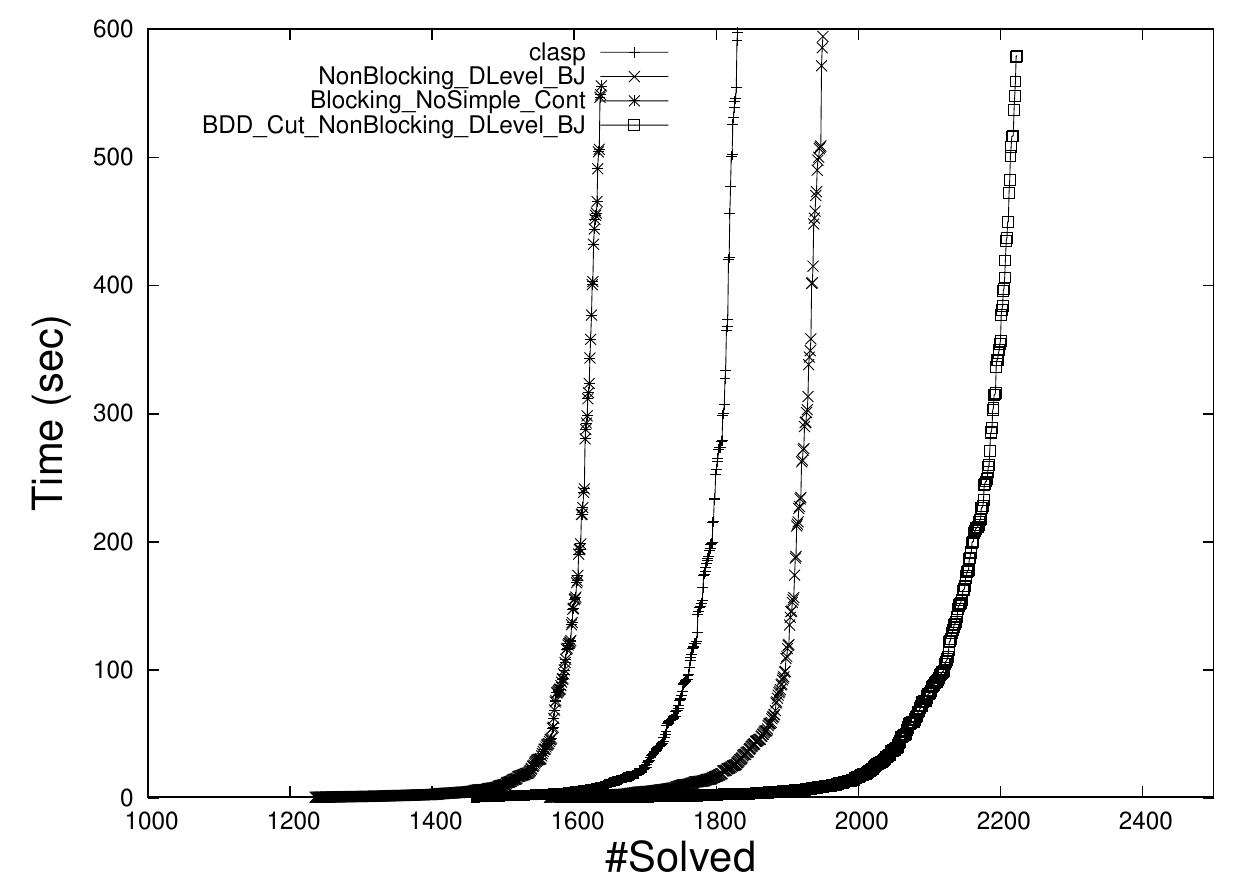}
\end{center}
\caption{Cactus plot of representative solvers with respect to running time.}\label{fig:cactus-all}
\end{figure*}

Figure~\ref{fig:cactus-all} shows a cactus plot of representative solvers.
Solved instances are ranked with respect to the times required to solve them.
Each point represents a solved instance with its rank (the horizontal coordinate) and the required time (the vertical coordinate).
Since one wants to solve as many instances as possible in a given amount of time, it is thought that gentler the slope of plotted points is, more efficient a solver is.
The formula-BDD caching solver clearly outperforms the other solvers.
It is then followed by the non-blocking solver, clasp, and the blocking solver in this order.

\begin{figure*}[t]
\begin{center}
\includegraphics[width=10cm]{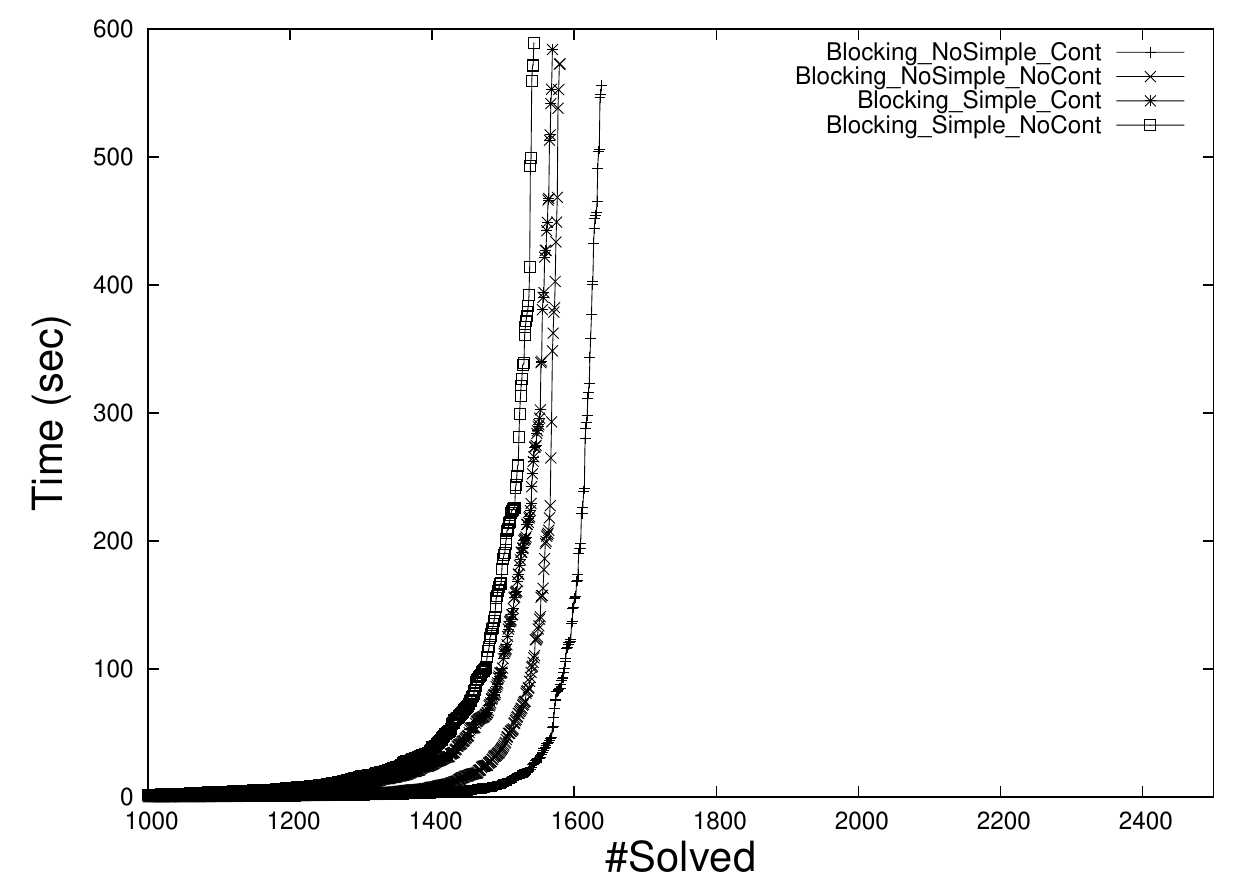}
\end{center}
\caption{Cactus plot of blocking solvers.}\label{fig:cactus-blocking}
\end{figure*}

\begin{figure*}[t]
\begin{center}
\includegraphics[width=10cm]{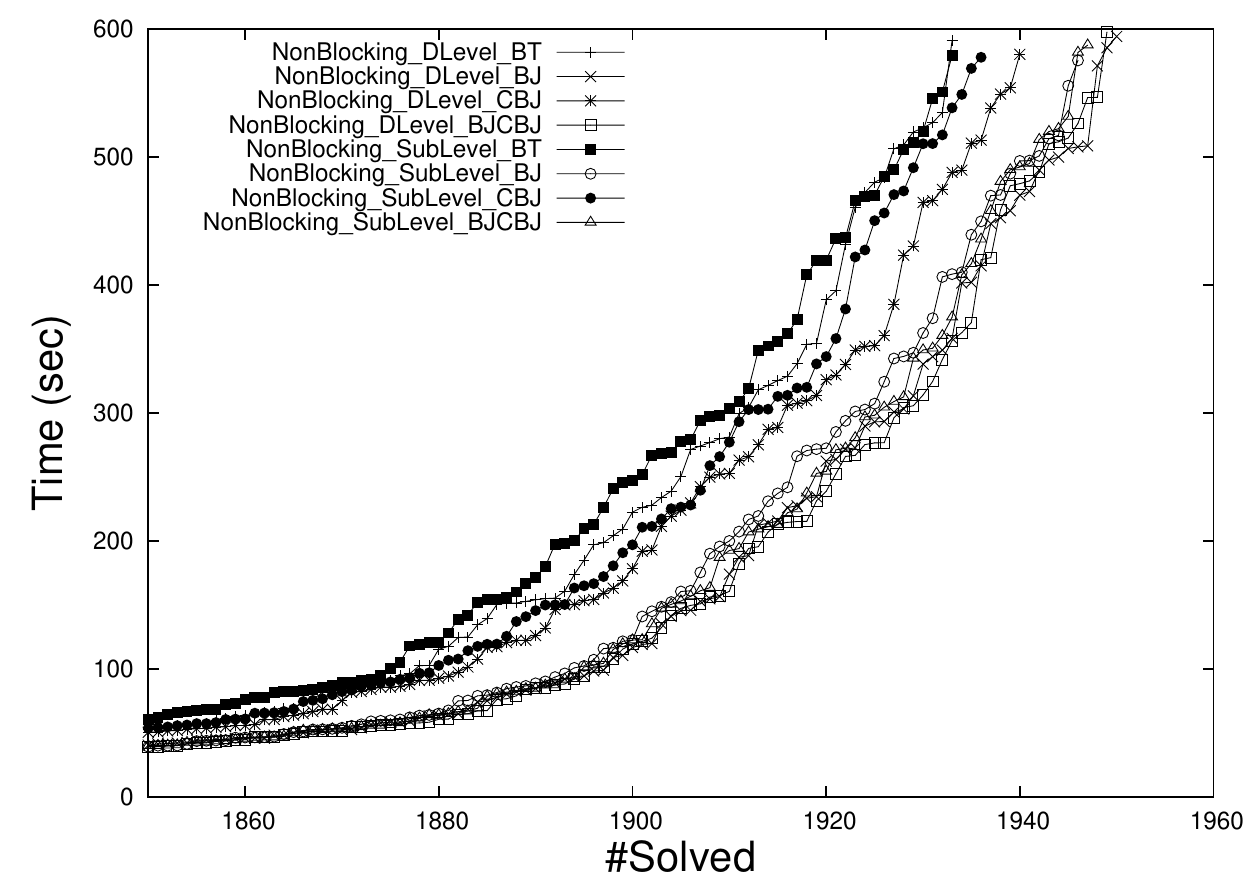}
\end{center}
\caption{Cactus plot of non-blocking solvers, where the horizontal scale is narrowed to make the difference clear.}\label{fig:cactus-nonblocking}
\end{figure*}

\begin{figure*}[t]
\begin{center}
\includegraphics[width=10cm]{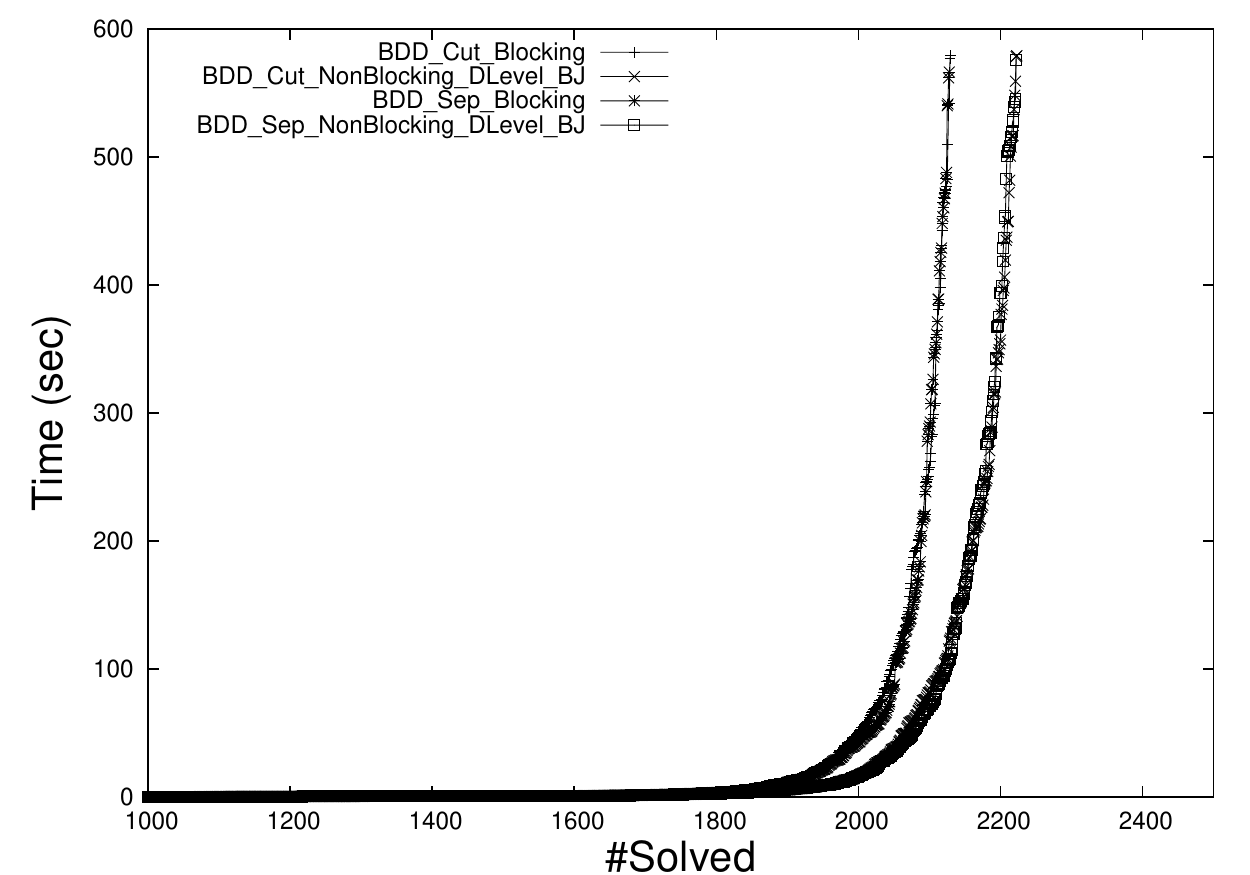}
\end{center}
\caption{Cactus plot of formula-BDD caching solvers.}\label{fig:cactus-bdd}
\end{figure*}

Figures~\ref{fig:cactus-blocking},~\ref{fig:cactus-nonblocking}, and~\ref{fig:cactus-bdd} depict differences between solvers of the same types.
From Figure~\ref{fig:cactus-blocking}, we can observe that continuation of search is effective yet simplification degrades performance.
For some instances, simplification enables a solver to find a large number of solutions, however such instances are limited and the current implementation is not powerful enough to make it possible to solve instances that can not be handled without simplification.
Figure~\ref{fig:cactus-nonblocking} has a narrower horizontal range than the other figures.
This is because non-blocking solvers exhibit quite similar performance and they can not be distinguished otherwise.
It is surprising that BT is almost as efficient as the other elaborated backtracking methods.
Decision level-based scheme is equal to or more efficient than sublevel-based scheme.
Figure~\ref{fig:cactus-bdd} shows that non-blocking procedure is clearly better as an underlying solver in which caching mechanism is embedded, while there is almost no difference between caching methods.

\subsection{Comparison of Maximum Memory Usage.}
\begin{figure*}[t]
\begin{center}
\includegraphics[width=10cm]{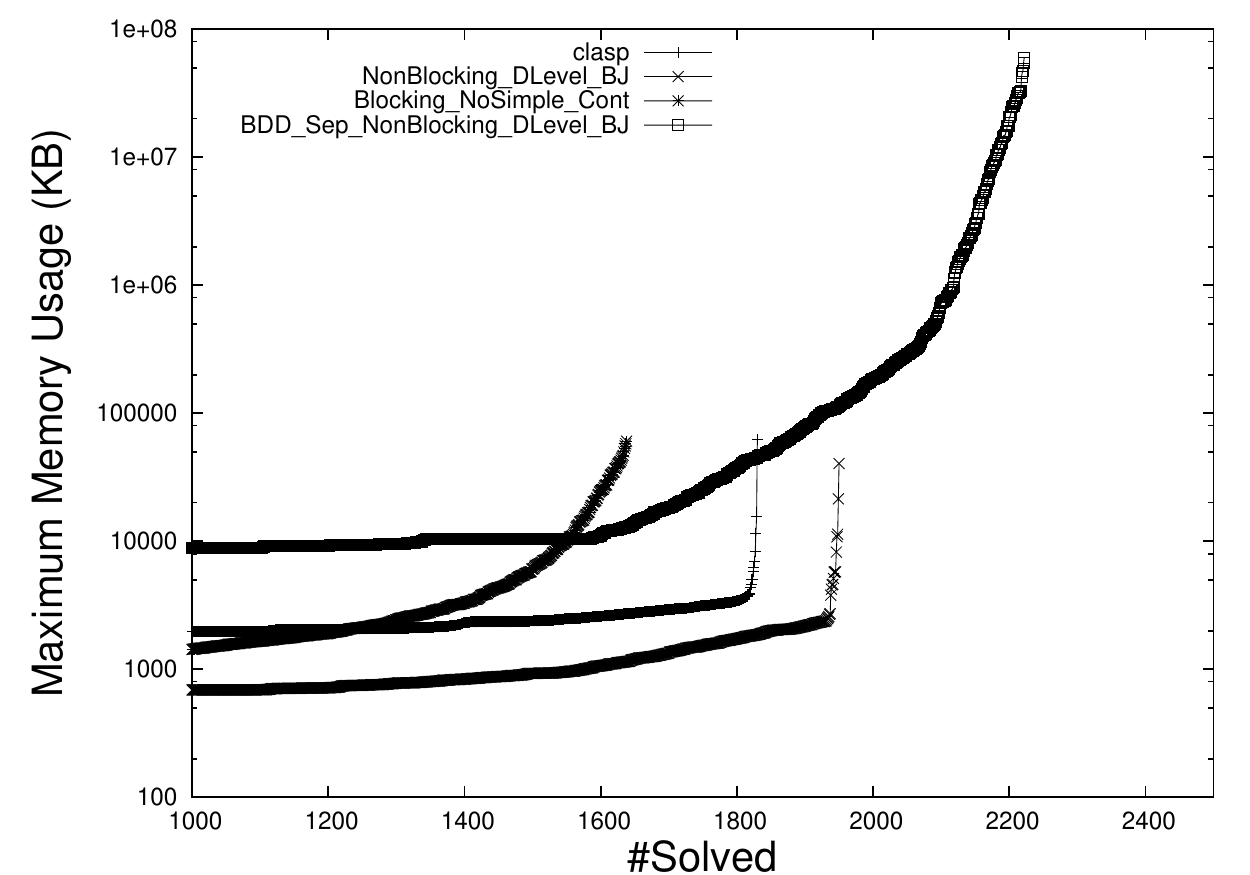}
\end{center}
\caption{Cactus plot of representative solvers with respect to the maximum memory usage, where the vertical axis is in logarithmic scale.}\label{fig:cactus-all-memory}
\end{figure*}

Figure~\ref{fig:cactus-all-memory} shows a cactus plot of the maximum memory usage.
Solved instances are in turn ranked with respect to the maximum memory usage.
Each point then represents a solved instance with its rank (the horizontal coordinate) and the required memory (the vertical coordinate).
In terms of memory consumption, the formula-BDD caching solver is the worst, while the non-blocking solver and clasp exhibit a stable performance.
The rapid increase in the curves of the non-blocking solver and clasp is due to large CNF formulae.
Although the formula-BDD caching solver consumes much memory, these days it is not unusual that even laptop computers have several giga bytes of RAM, and advantage of the formula-BDD caching solver is not impaired so much.

\subsection{Comparison of Scalability in The Number of Solutions.}\label{subsec:scalability}
As shown in Table~\ref{tab:dist1}, each representative solver has the following limit in the number of solutions within 600 seconds time limit .
\begin{itemize}
\item \textbf{Blocking}:     one million solutions.
\item \textbf{Clasp}:        one hundred million solutions.
\item \textbf{Non-Blocking}: ten billion solutions.
\item \textbf{BDD}:          more than one quadrillion solutions.
\end{itemize}

\begin{table}[t]
    \begin{center}
    \caption{Distribution of solved instances with respect to the number of solutions.\label{tab:dist1}}
        \begin{tabular}{l|rrrr}
                                  & Blocking & clasp & NonBlocking & BDD \\ \hline
       $\intvlr{0}{10^1}$         & 38       & 36    & 39          & 37  \\
       $\intvr{10^1}{10^2}$       & 11       & 11    & 11          & 11  \\
       $\intvr{10^2}{10^3}$       & 417      & 417   & 417         & 416 \\
       $\intvr{10^3}{10^4}$       & 682      & 682   & 682         & 682 \\
       $\intvr{10^4}{10^5}$       & 408      & 408   & 408         & 408 \\
       $\intvr{10^5}{10^6}$       & 82       & 134   & 133         & 134 \\
       $\intvr{10^6}{10^7}$       & 0        & 86    & 82          & 86  \\
       $\intvr{10^7}{10^8}$       & 0        & 56    & 87          & 91  \\
       $\intvr{10^8}{10^9}$       & 0        & 0     & 68          & 83  \\
       $\intvr{10^9}{10^{10}}$    & 0        & 0     & 23          & 92  \\
       $\intvr{10^{10}}{10^{11}}$ & 0        & 0     & 0           & 83  \\
       $\intvr{10^{11}}{10^{12}}$ & 0        & 0     & 0           & 44  \\
       $\intvr{10^{12}}{10^{13}}$ & 0        & 0     & 0           & 29  \\
       $\intvr{10^{13}}{10^{14}}$ & 0        & 0     & 0           & 19  \\
       $\intvr{10^{14}}{\infty}$  & 0        & 0     & 0           & 8   \\ \hline
        total                     & 1,638    & 1,830 & 1,950       & 2,223
        \end{tabular}
    \end{center}
\end{table}

\subsection{Distribution of Solved Instances over Instance Series.}\label{subsec:solvedinstances}
Table~\ref{tab:dist3} shows the distribution of solved instances over all instance series.
In almost all series, differences in the number of solved instances between solvers can be explained by their scalability.

Exceptions are the \textbf{sc14} instances.
They are clearly harder instances.
All solvers were unable to find even one solution in many instances.
For \textbf{sc14-crafted} series, solved instances are those of less than 10 solutions with a few exceptions of BDD.
Table~\ref{tab:dist4} shows the distribution of all \textbf{sc14} instances including unsolved instances.
Although clasp could find relatively many solutions, the other solvers could only find less than 10 solutions for a majority of instances.
In terms of the ability to find solutions for \textbf{sc14} instances, we could say that the best solver is clasp; the blocking solvers and the non-blocking solvers are a tie, both ranked at the second; the formula-BDD caching solver is no match for those instances, which is due to fixed variable ordering.

\begin{figure*}[t]
\begin{center}
\includegraphics[width=7cm]{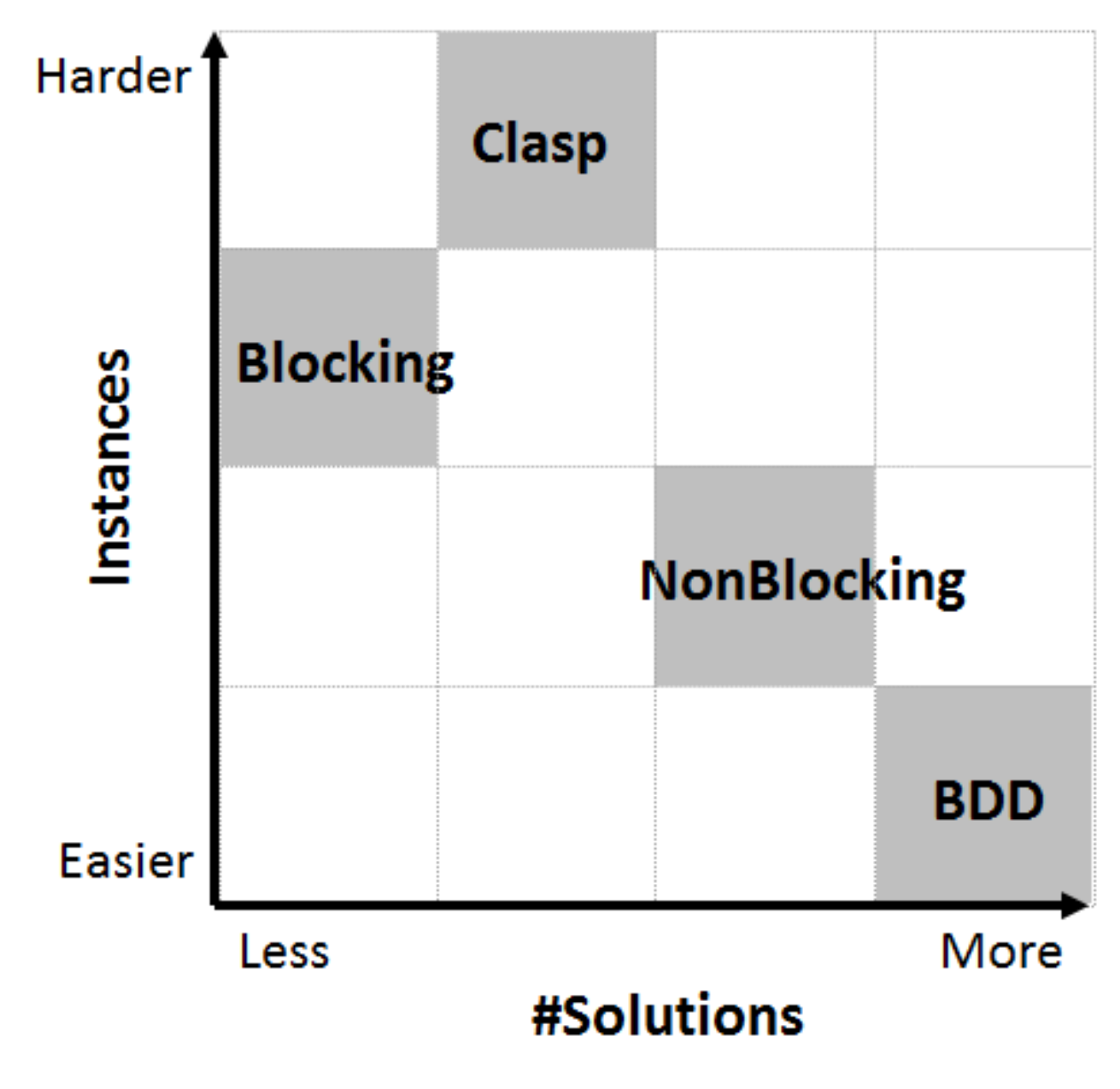}
\end{center}
\caption{Favorite ranges of instances for representative solvers.}\label{fig:map}
\end{figure*}

The favorite ranges of instances for representative solvers are illustrated in Fig.~\ref{fig:map} according to the two factors: hardness of instances and the numbers of solutions instances have.
Each solver is placed in such a way that the vertical position corresponds to the ability to find solutions for \textbf{sc14} instances, where the blocking solver is above the non-blocking solver because of more benefit from a SAT solver, and the horizontal position corresponds to the scalability in the number of solutions.
Hence, each indicated range refers to instances leftward or downward from it as well as those within it.
It should be noted that shrinking the vertical axis would be more suitable for real performance, because in reality all solvers, on the whole, exhibits a poor performance over hard instances.

\begin{table}[t]
    \begin{center}
    \caption{Distribution of solved instances over instance series, where the number of instances in each series is enclosed in parenthesis.\label{tab:dist3}}
        \begin{tabular}{lr|rrrr}
                            &        & Blocking & clasp & NonBlocking & BDD \\ \hline
        ais                 & (4)    & 4        & 4     & 4           & 4   \\
        bmc                 & (13)   & 0        & 0     & 0           & 1   \\
        bw                  & (7)    & 7        & 7     & 7           & 6   \\
        Flat125-301         & (100)  & 4        & 51    & 94          & 100 \\
        Flat150-360         & (101)  & 0        & 12    & 61          & 101 \\
        Flat175-417         & (100)  & 0        & 2     & 15          & 98  \\
        Flat200-479         & (100)  & 0        & 0     & 3           & 77  \\
        Flat75-180          & (100)  & 87       & 100   & 100         & 100 \\
        flat                & (1,199)& 1,136    & 1,195 & 1,199       & 1,199\\
        gcp                 & (1)    & 0        & 0     & 0           & 0   \\
        hanoi               & (2)    & 2        & 2     & 2           & 2   \\
        inductive           & (41)   & 1        & 5     & 9           & 8   \\
        logistics           & (4)    & 0        & 0     & 0           & 1   \\
        parity              & (20)   & 20       & 20    & 20          & 20  \\
        qg                  & (10)   & 10       & 10    & 10          & 10  \\
        ssa                 & (4)    & 0        & 0     & 0           & 0   \\
        SW100-8-0           & (100)  & 0        & 0     & 0           & 0   \\
        SW100-8-1           & (100)  & 0        & 0     & 0           & 0   \\
        SW100-8-2           & (100)  & 0        & 0     & 0           & 0   \\
        SW100-8-3           & (100)  & 0        & 0     & 0           & 0   \\
        SW100-8-4           & (100)  & 0        & 1     & 5           & 68  \\
        SW100-8-5           & (100)  & 55       & 100   & 95          & 100 \\
        SW100-8-6           & (100)  & 100      & 100   & 100         & 100 \\
        SW100-8-7           & (100)  & 100      & 100   & 100         & 100 \\
        SW100-8-8           & (100)  & 100      & 100   & 100         & 100 \\
        SW100-8-p0          & (1)    & 1        & 1     & 1           & 1   \\
        sc14-app            & (56)   & 0        & 0     & 0           & 0   \\
        sc14-crafted        & (65)   & 5        & 3     & 6           & 5   \\
        iscas               & (39)   & 6        & 17    & 19          & 22  \\ \hline
        total               & (2,867)& 1,638    & 1,830 & 1,950       & 2,223
        \end{tabular}
    \end{center}
\end{table}

\begin{table}[t]
    \begin{center}
    \caption{Distribution of all sc14 instances including unsolved instances according to the number of found solutions within time limit.\label{tab:dist4}}
        \begin{tabular}{l|rrrr}
                                  & Blocking & clasp & NonBlocking & BDD \\ \hline
       $\intv{0}{0}$              & 54       & 22    & 58          & 89  \\
       $\intvlr{0}{10^1}$         & 25       & 24    & 22          & 19  \\
       $\intvr{10^1}{10^2}$       & 8        & 0     & 0           & 0   \\
       $\intvr{10^2}{10^3}$       & 12       & 1     & 0           & 0   \\
       $\intvr{10^3}{10^4}$       & 3        & 5     & 1           & 0   \\
       $\intvr{10^4}{10^5}$       & 4        & 20    & 1           & 0   \\
       $\intvr{10^5}{10^6}$       & 15       & 27    & 2           & 0   \\
       $\intvr{10^6}{10^7}$       & 0        & 21    & 0           & 2   \\
       $\intvr{10^7}{10^8}$       & 0        & 1     & 0           & 0   \\
       $\intvr{10^8}{10^9}$       & 0        & 0     & 14          & 0   \\
       $\intvr{10^9}{10^{10}}$    & 0        & 0     & 23          & 0   \\
       $\intvr{10^{10}}{10^{11}}$ & 0        & 0     & 0           & 0   \\
       $\intvr{10^{11}}{10^{12}}$ & 0        & 0     & 0           & 1   \\
       $\intvr{10^{12}}{10^{13}}$ & 0        & 0     & 0           & 1   \\
       $\intvr{10^{13}}{10^{14}}$ & 0        & 0     & 0           & 0   \\
       $\intvr{10^{14}}{\infty}$  & 0        & 0     & 0           & 9   \\ \hline
        total                     & 121      & 121   & 121         & 121
        \end{tabular}
    \end{center}
\end{table}

\subsection{Comparisons to Publicly Available \#SAT Solvers}\label{subsec:sharpsat}
We also have a comparison with publicly available \#SAT solvers, sharpSAT version 1.1~\cite{978-3-540-37206-6}, c2d version 2.20~\cite{DBLP:conf/ecai/Darwiche04} and relsat version 2.20~\cite{Bayardo:1997:UCL:1867406.1867438}, to check the performance of our developed ALLSAT solvers for \#SAT.
In the result for the same 2867 instances, relsat, sharpSAT, c2d and BDD\_Cut\_NonBlocking\_DLevel\_BJ solved 2191, 2196, 2196 and 2223 respectively.
Since relsat supports solution count, it was used as a solution counter in this comparison.
So, even for \#SAT, the formula-BDD caching solver shows better performance. 

\section{Conclusion}\label{sec:conc}
We surveyed and discussed major techniques of existing AllSAT solvers.
We classified the types of solvers into a blocking solver, a non-blocking solver, and a formula-BDD caching solver.
We faithfully implemented and released these solvers publicly so that other researchers can easily develop their solver by modifying our codes and compare it with existing methods.
We conducted comprehensive experiments with total 2867 instances taken from SATLIB, SAT competition 2014, and ISCAS benchmarks.
Apart from our implemented solvers, we used clasp, one of the few off-the-shelf softwares with solution generation support.
The experiments revealed the following solver's characteristics (600 seconds time limit). See also Fig.~\ref{fig:map}.
\begin{itemize}
\item The formula-BDD caching solver is the most powerful.
It has the most solved instances, including instances with more than one quadrillion solutions.
The maximum memory usage amounts to several tens giga bytes in the worst case, though it is controllable by refreshing caches at the cost of a low cache hit rate.
They are bad at hard instances due to fixed variable ordering.
\item The non-blocking solver is ranked at the next best, followed by clasp.
The non-blocking solver and clasp can handle instances with ten billion solutions and one hundred million solutions with a low maximum memory usage (a few mega bytes to several tens of mega bytes), respectively.
Although both solvers exhibit relatively a similar performance, a difference is that clasp is able to find a moderately many number of solutions from even hard instances, though it is not powerful enough to make it possible to solve instances that can not be handled by other means.
\item The blocking solver is limited to instances with one million solutions as blocking clauses deteriorate the performance of unit propagation.
However, it can benefit from state-of-the-art techniques of SAT solvers as they are, thereby it is suitable for finding a small number of solutions for hard instances.
\end{itemize}

From the above, we conclude that the formula-BDD caching solver is the most superior in terms of exact AllSAT solving over various kinds of instances.
However, since not all solutions are necessary in some practical applications and duplicated solutions may be allowed, it is recommended to select an appropriate solver in accordance with types of instances and applications.

\bibliographystyle{plain}
\bibliography{manuscript} 

\end{document}